\documentclass[12pt]{article}
\usepackage{amsmath}
\usepackage{amssymb}
\usepackage{bm}
\usepackage{graphicx,psfrag,epsf}
\usepackage{array}  
\usepackage{enumerate}
\usepackage{natbib}
\usepackage{url} 
\usepackage{dsfont}
\usepackage{tikz}
\usetikzlibrary{shapes, arrows.meta, positioning}
\usepackage{makecell}

\newcommand{\PP}{{\mathbb{P}}}

\newcommand{\bY}{\bm{Y}}
\newcommand{\bX}{\bm{X}}

\newcommand{\ellin}{\ell \in \{E,C\}}

\newcommand{\NTB}{\Delta_{\bm\tau}}

\newcommand{\blind}{0}

\begin{document}

\def\spacingset#1{\renewcommand{\baselinestretch}%
{#1}\small\normalsize} \spacingset{1}


\if0\blind
{
  \title{\bf Navigating the Landscape of Hierarchical Multi-Component Strategies: \\GPC, DOOR, and MOST}

  \author{%
    Micka\"el De Backer$^{1}$ \and
    Johan Verbeeck$^{2}$ \and
    Vivian Lanius$^{3}$ \and
    Marc Vandemeulebroecke$^{4}$ \and
    Scott Evans$^{5,6}$ \and
    Toshimitsu Hamasaki$^{5,6}$ \and
    Marc Buyse$^{2,7,8}$ \and
    Frank E. Harrell Jr.$^{9}$
  }

  \date{}
  \maketitle

  \begingroup\footnotesize
  \noindent$^{1}$UCB, Anderlecht, Belgium. \texttt{mickael.debacker@ucb.com}\\
  $^{2}$Data Science Institute, I-BioStat, Hasselt University, Diepenbeek, Belgium\\
  $^{3}$UCB, Monheim am Rhein, Germany\\
  $^{4}$Clinical Statistics \& Analytics, Statistical Innovation, Bayer BCC AG, Basel, Switzerland\\
  $^{5}$The Biostatistics Center, Milken Institute School of Public Health, The George Washington University, Washington, DC, USA \\
  $^{6}$Department of Biostatistics and Bioinformatics, Milken Institute School of Public Health, George Washington University, Washington, DC, USA\\
  $^{7}$Department of Biostatistics, IDDI (International Drug Development Institute), Ottignies, Belgium\\
  $^{8}$One2Treat, Louvain-la-Neuve, Belgium \\
  $^{9}$Department of Biostatistics, Vanderbilt University School of 
Medicine, Nashville, TN, USA
  \par\endgroup

} \fi

\bigskip
\begin{abstract}
There is a growing recognition of the importance to involve patients in every stage of drug development. This shift acknowledges that patients' perspectives, experiences, and preferences are essential for ensuring that treatments meet real-world needs. In this context, a new body of statistical literature has emerged, focusing not only on the simultaneous consideration of multiple outcomes that reflect patients' overall experiences, but also on their structured prioritization. We refer to this class of approaches as hierarchical multi-component statistical methods. Among these, two influential frameworks - generalized pairwise comparisons (GPC) and desirability of outcome ranking (DOOR) - have emerged in the last decade, each aiming to offer a comprehensive approach to evaluating treatment effects. A new methodology, referred to here as the Markov ordinal state transition model (MOST), has recently been introduced without focusing on an explicit link with GPC nor DOOR. This paper seeks to fill this gap by offering a comprehensive and comparative analysis of the three approaches. Through examples and an exploration of the structural and philosophical differences between the methods, our aim is to provide guidance and encourage lines of research in the rapidly-evolving landscape of hierarchical multi-component statistical methodologies. 
\end{abstract}

\noindent%
{\it Keywords:} multiple outcomes; ordinal analysis; longitudinal analysis; clinical desirability
\vfill

\newpage
\spacingset{1.45} 


\section{Introduction}
Randomized clinical trials have traditionally focused on a single primary outcome, chosen as the key measure of treatment efficacy and aligned with good practice by reflecting how a patient survives, feels, and functions. Ideally, this outcome would fully reflect the nature of the disease, enable a clear and comprehensive interpretation of the intervention's effect, while permitting a statistically-efficient detection of the `treatment effect.' However, in many clinical settings, such an ideal outcome does not exist, as patient experiences are inherently multidimensional due to the multiple facets of the disease.

A more comprehensive evaluation of a treatment may thus come with the analysis of several outcomes of interest. Traditionally, and out of convenience, this has been done following the process of analyzing first and then aggregating: outcomes are considered univariately and analyzed separately, with their results then evaluated and communicated at an aggregated level. This approach, however, has two major limitations as pointed out for instance in \citet{evans2022our} and \citet{hamasaki2025patient}. First, it may fail to accumulate evidence of the benefits of the intervention when individual benefits are challenging to demonstrate separately within the constraints of a typical randomized trial. A separate analysis of several key characteristics of the disease may then fall short to detect those benefits, while aggregating them in a clinically-sensible manner into a single analysis may help strengthen the signal. This is the usual motivation for resorting to composite endpoints in clinical trials, although the latter often struggle to combine outcomes in a way that truly reflects clinical priorities (\textit{e.g.}, by assigning equal importance to death and progression in the endpoint progression-free survival in oncology).

The second limitation arising from conducting separate analyses is the failure to capture clinically-important associations between outcomes. For instance, in benefit-risk evaluations, univariate analyses cannot determine whether severe side effects occur in the same patients who benefit from the experimental treatment in terms of efficacy, or if the opposite is true. Associations also play a crucial role in contexts involving intercurrent events that influence the interpretation of the primary outcome. For instance, in hidradenitis suppurativa (chronic, painful skin inflammation), the interpretation of the number of inflammatory nodules observed at a given time is inherently influenced by the use and amount of rescue medication; our clinical understanding of a value on one outcome is thus conditional on knowledge from other measurements. Incorporating such nuance, by integrating intercurrent events into the outcome of interest, corresponds to the so-called composite strategy of the estimand framework discussed in ICHE9(R1) (\citet{ICHE9}).

In recent years, several statistical methodologies have been developed to address the challenges outlined above and to enable joint analyses that would supplement traditional univariate approaches. Two of the main methods, particularly used in late-phase trials, are generalized pairwise comparisons (GPC, \cite{buyse2010generalized} and \cite{pocock2012win}) and desirability of outcome ranking (DOOR, \cite{evans2015desirability}). What fundamentally unites these methods is the ambition to integrate outcomes of different types into a single statistical analysis, guided by clinical considerations. Typically, these clinical considerations are incorporated through a concept of ordering: an order of clinical desirability for outcomes is established, requiring only the identification of what is more desirable without the need for (explicitly) assigning weights to each of the outcomes. This ordering makes the methods distinct from some traditional composite endpoints (\textit{e.g.}, time-to-first event endpoints) and other attempts at multivariate analyses (\textit{e.g.}, joint models as in \citet{ivanova2016mixed} or \citet{iddi2012combined}). Accordingly, this paper focuses on contexts where such an ordering is meaningful. In this respect, the methods discussed here represent an extension to established multi-component endpoints, such as the modified Rankin Scale (mRS) in stroke or the Expanded Disability Status Scale (EDSS) in multiple sclerosis, which, though developed earlier, share the same underlying principles.

GPC and DOOR are often presented with a twofold promise echoing the challenges evoked previously: i) to provide a statistically-sensitive \textit{and} clinically-relevant evaluation of an experimental treatment's effect in contrast to the control, and ii) to improve insight into treatment effects when multiple outcomes are considered (\cite{buysemolenberghs2025gpc}, \cite{evans2015desirability}). However, the methods differ in how they aim to achieve these goals, and as a result, they have developed in parallel with limited overlap. 

Based on the considerations outlined above, both GPC and DOOR have gained traction in practice. GPC, in particular, has made significant advances in cardiovascular studies (see \citet{pocock2024win} for a recent overview of the Win Ratio, one of the GPC-based statistics), as well as in kidney diseases (see \textit{e.g.}, \citet{little2023validity}). Other key areas where GPC is being applied include oncology and rare diseases (see, for example, \citet{saadapplications} and \citet{deltuvaiteapplications}). In parallel, DOOR has traditionally been used in infectious diseases, though its application is expanding into other fields such as neurology and obstetrics, among others (see \citet{Toshimitsuapplications} for a comprehensive overview). Both GPC and DOOR are also used in benefit-risk evaluations to assess whether the risk-benefit balance of patients is positive for the experimental treatment in contrast to the control, which in some cases can advantageously circumvent testing for conventional non-inferiority (see, for instance, \citet{evans2015desirability}, \citet{backer2024design}, and \citet{verbeeck2025non}).

Despite their advantages and increasing use, GPC and DOOR also raise important challenges, statistical as well as communicative, including those associated with aligning their use to the estimand framework. These points will be developed in Sections \ref{sec:philo_diff} and \ref{sec:stat_diff}. Meanwhile, a new methodology - Markov ordinal state transition model (MOST) - has recently emerged (see, \textit{e.g.}, \citet{rohde2024bayesian} for a tutorial on applying this approach). By its very nature, MOST addresses several of the challenges faced by DOOR and GPC as will be detailed in this paper. At first glance, and as will be explained later, MOST in fact appears to be a longitudinal version of DOOR. However, there is more to this approach, offering potential improvements both in terms of statistical properties and in the way results can be communicated and interpreted. These points will be explored in detail throughout this paper.

It is the objective of this paper to provide a general overview of GPC, DOOR and MOST, with an emphasis on their distinct characteristics. To structure the paper, we will make a distinction between differences in philosophical characteristics, and differences in statistical characteristics. While the paper does not focus on new methodological developments, it offers foundational perspectives that can help guide the appropriate application of each approach. These concepts will be illustrated using a single guiding example carried throughout the paper. More broadly, our objective is to encourage and promote further research into hierarchical multi-component statistical methods. Ultimately, we hope to help addressing the ongoing challenge that many clinical trials fail to generate the evidence needed to support effective medical decision-making (\citet{demets2011historical}).

This paper is structured as follows: Section \ref{sec:guiding_example} introduces a guiding example from critical care medicine, including a description of the fictitious dataset (though inspired by a real randomized trial) and an overview of conventional methods of analysis. Section \ref{sec:pf_stat_meth} presents the GPC, DOOR, and MOST methodologies. Each is first described in general terms, followed by a brief application to the guiding example. Sections \ref{sec:philo_diff} and \ref{sec:stat_diff} form the core of the paper, providing a point-by-point comparison of the key philosophical and statistical differences between the methods, respectively. Section \ref{sec:further_research} outlines potential directions for future research, and Section \ref{sec:discussion} concludes with a general discussion.


\section{A guiding example}\label{sec:guiding_example}
In this section, we present an example that will be used to contrast the three methodologies of interest. This example is not per se designed to offer a fair comparison in terms of statistical properties. Rather, its purpose is to illustrate the nature of each methodology and highlight some key differences in their underlying principles. In that sense, we will first introduce the example before analyzing the latter in four different and independent ways, using either conventional statistical tools or one of the three methodologies of interest in this paper.

\subsection{Context}
We consider a fictitious example of clinical trial data in critical care medicine, using material available in \citet{fharrellSim}. A detailed explanation of the underlying mechanism for simulating the data is available in the Supplementary Material. 

The data are representing a study involving 200 patients, randomized in a 1:1 ratio to either experimental treatment or control. The recorded data includes the start and end dates of hospitalization, discharge, transfers between wards, the date of death if applicable, and the dates when ventilation was initiated and ended. This can be translated into a daily health status over 28 days from baseline for each patient, classified into one of four categories: at home, in hospital, on ventilator, or dead. At baseline (day 0), all patients are hospitalized. For simplicity, we assume complete data with no missing observations throughout the study, although we will comment on the impact of missing data in Section \ref{sec:stat_diff}. 

\begin{figure}[t!]
\centering
\includegraphics[width=0.49\linewidth]{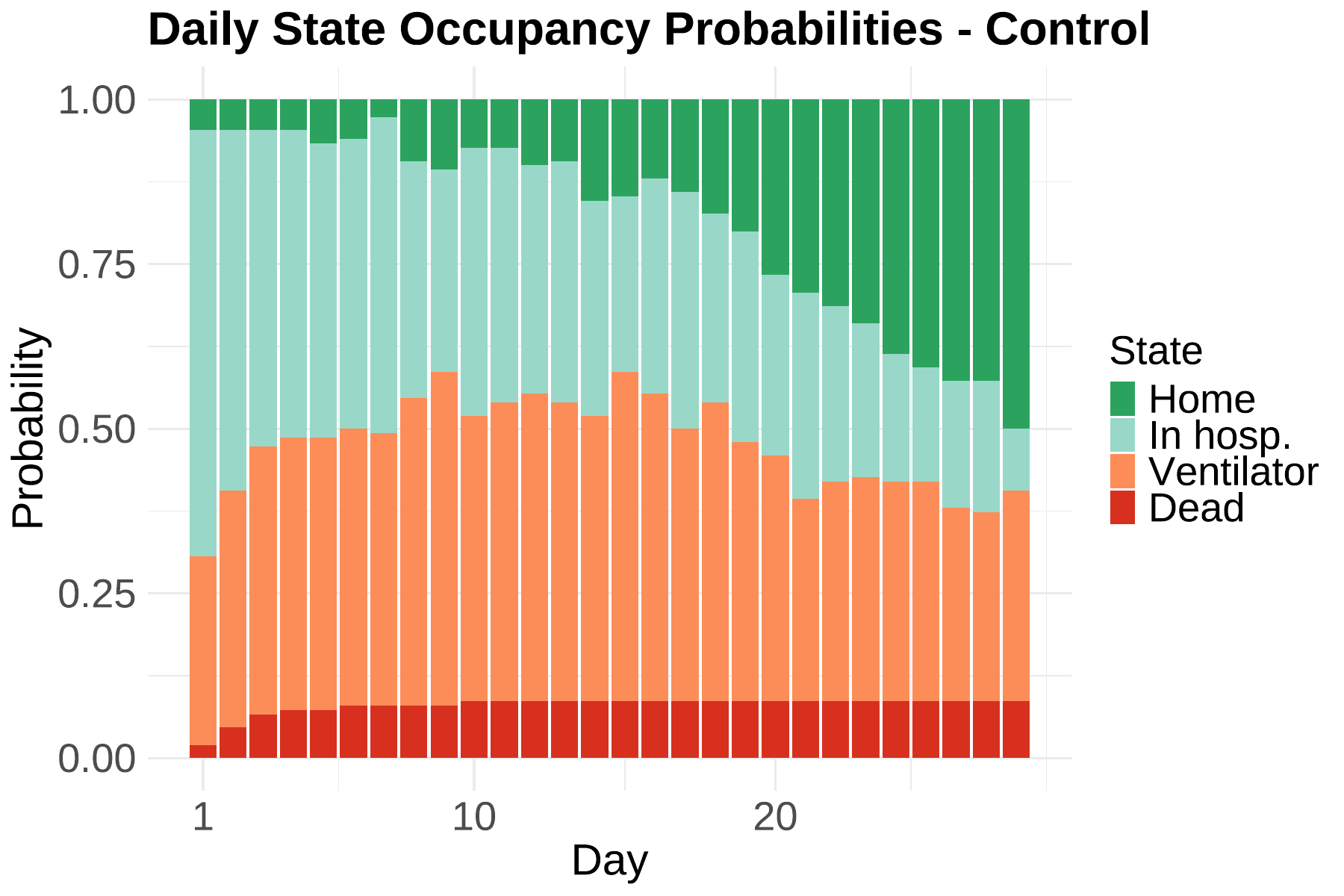}
\includegraphics[width=0.49\linewidth]{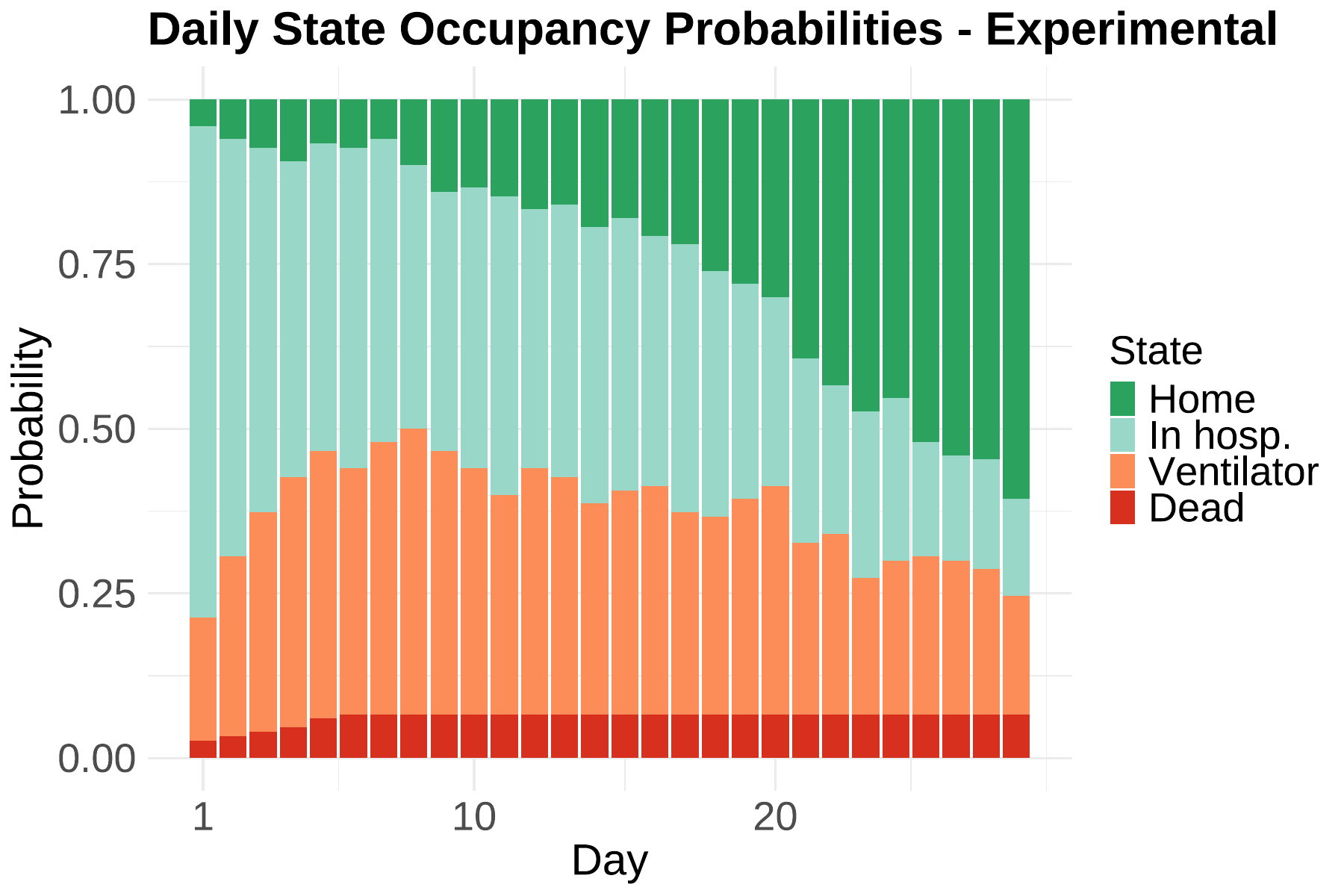}
\caption{\label{fig:sop_guiding_example} Empirical state occupancy probabilities of patients under control and experimental treatment for the guiding example.}
\end{figure}

For illustration, Figure \ref{fig:sop_guiding_example} displays the empirical state occupancy probabilities for each arm, that is, the empirical probability of patients being in each of the four health states at each time point. Several patterns emerge from the plot: minimal differences in mortality between the two arms, a general trend of health improvement in the second part of the trial even in the control arm, and a progressive benefit associated with the experimental treatment over time. Since death is an absorbing state, the proportion of patients in that state is non-decreasing over time.

\subsection{An Option for a Conventional Analysis}
We start with an analysis of the data that does not build upon the methodologies of interest in this paper. The analysis would then typically make a choice for a primary outcome, focusing in this example on either time-to-recovery (with context-specific definitions of `recovery', see \textit{e.g.}, \citet{beigel2020remdesivir}) or the number of ventilator-free days. We concentrate here on the latter, as time-to-recovery is i) prone to the competing risk of death (which complicates interpretation and fails to penalize efficacy for death), and ii) it ignores the possibility for some patients to relapse. Number of ventilator-free days is also affected by death, but this is usually handled by creating an overriding category with value $<0$ for that outcome in case of death. 

Formally, we let $Y_i^\ell$, $i=1,\ldots,n_\ell$, $\ell \in \{E, C\}$, be the i.i.d. ventilator-free days where the superscripts $E$ and $C$ are used to denote the experimental treatment and control, respectively. In the absence of adjusting for baseline characteristics (a topic that will be evoked in Section \ref{sec:stat_diff}), a conventional non-parametric analysis of the now ordinal endpoint (including the overriding category for death) could be based on the Wilcoxon-Mann-Whitney test (WMW, see \citet{wilcoxon}, \citet{mann1947test}). We focus here on its U-statistic version in contrast to its rank-based version. In particular, the test is based on the comparison of each individual in the experimental arm with every individual in the control arm. The comparison of each pair of patients is then assigned a score $U_{ij}$, with 
$$
U_{ij} = \left[Y_i^E > Y_j^C \right] - \left [Y_i^E < Y_j^C \right ],
$$ 
for $i=1,\ldots,n_E$, $j=1, \ldots, n_C$, and where $[\cdot]$ denotes the indicator function, taking value 1 if the event is true and 0 otherwise. The score is here written assuming that higher values of $Y$ are more desirable, and it uses a slightly different convention from the original formulation of the WMW test in order to anticipate an evident link with GPC. 

In essence, each pairwise comparison receives a value of either 1 (the contrast is in favor of the patient in the experimental arm), -1 (the contrast is in favor of the patient in the control arm) or 0 (the contrast does not favor any patient). Assembling all the pairwise comparisons, the quantity $(n_E n_C)^{-1}\sum_{i=1}^{n^E}\sum_{j=1}^{n^C} U_{ij}$ then forms the basis of the analysis. Treatment effects are then quantified through the limiting value of the averages of pairwise comparisons, \textit{i.e.}, the probabilities\footnote{We keep here the subscripts $i$ and $j$ to insist on the fact that the contrast is for two \textit{distinct} patients - which is different from a contrast of potential outcomes for the same patient under either the experimental treatment or the control, see \textit{e.g.}, \citet{fay2018causal}.} $\mathbb{P}(Y^E_i > Y^C_j)$, $\mathbb{P}(Y^E_i < Y^C_j)$ and $\mathbb{P}(Y^E_i = Y^C_j)$. For instance, absolute and relative measures of treatment effect are given, respectively, by the so-called Net Treatment Benefit (NTB; $\Delta$, \citet{buyse2021net}) and the Win Odds ($\Theta$, \citet{dong2020win}), defined as:
\[
\Delta := \PP(Y^E_i > Y^C_j) - \PP(Y^E_i < Y^C_j), \qquad \Theta := \frac{\PP(Y^E_i > Y^C_j) + 1/2 \, \PP(Y^E_i = Y^C_j)}{\PP(Y^E_i < Y^C_j) + 1/2 \, \PP(Y^E_i = Y^C_j)}.
\]
The Win Ratio ($\Psi$), commonly used in cardiovascular studies, is a relative measure of treatment effect as well and is defined as $\Psi := \PP(Y_i^E > Y_j^C) / \PP(Y_i^E < Y_j^C)$. However, we do not consider the WR further in this paper because it does not properly account for situations where no clear winner can be determined between individuals (see, \textit{e.g.}, \citet{brunner2021win} and \citet{verbeeck2023generalized}). We will use the approach of the WMW test described above to build on the hierarchical multi-component methodologies in the next section.


\section{The hierarchical multi-component methodologies}\label{sec:pf_stat_meth}
We present in this section the three hierarchical multi-component methodologies of interest. We focus here mainly on how one would independently apply these methodologies to the guiding example, before elaborating on their core philosophical differences in the next section. 

\subsection{Generalized pairwise comparisons}
 
We start with GPC, as it is a direct extension of the conventional analysis described earlier. Specifically, GPC follows a similar approach to the WMW test, relying on the comparison of pairs of individuals across different arms to determine whether one individual has a more favorable outcome than the other in the pair. The advancement of GPC over the WMW test is its broader definition of ``more favorable.'' Two main extensions are introduced: i) for univariate outcomes, GPC allows to recognize that not all numerical differences are clinically meaningful, and ii) the definition of a ``more favorable'' outcome can be multivariate, allowing for the incorporation of various aspects of the effect of the experimental treatment. We use the guiding example to demonstrate these features. A detailed description of recent advances in the GPC literature can be found in \citet{buyse2025gpc}.

Consider the evaluation of ventilator-free days, with two pairs of observed values: $(Y^E_i, Y^C_j) = (27, 2)$ and $(Y^E_k, Y^C_j) = (3, 2)$. In the WMW test, both pairs would receive a score of +1, since in each case, the patient in the experimental arm has more ventilator-free days than the patient in the control arm. In contrast, GPC introduces a more nuanced approach to assessing the clinical relevance of numerical differences: for instance, one might decide that in the second pair, where the difference is only 1 day, there is no substantial ``winner'' in the comparison. The contrast between pairs is therefore adjusted based on clinical considerations, which can vary depending on the context. Formally, one shifts from the scores $U_{ij}$ towards 
$
U_{ij}^\tau = \left [Y_i^E \succ_\tau Y_j^C\right] -\left [Y_i^E \prec_\tau Y_j^C \right],
$
where $\succ_\tau$ denotes a `more favorable result accounting for clinical considerations', and similarly for $\prec_\tau$. In practice, the contrast is often considered in the form of $\left [Y_i^E \succ_\tau Y_j^C\right] = \left [Y_i^E > Y_j^C + \tau\right]$ with $\tau > 0$, although the idea is broader than that\footnote{This approach reveals an interesting, almost paradoxical ambition: it introduces a notion of scale via the parameter $\tau$ into the Wilcoxon–Mann–Whitney (WMW) test, a method traditionally valued for its scale-invariance. This is not without challenges: see \textit{e.g.}, \citet{burzykowskilimitations}.}. Using this formulation of clinically-relevant contrasts, a GPC-based analysis then builds upon aggregation of the scores $ U_{ij}^\tau$, with corresponding absolute or relative treatment effects of the form:
\[ \Delta_\tau := \PP(Y^E_i > Y^C_j + \tau) - \PP(Y^E_i + \tau < Y^C_j), \; \Theta_\tau := \frac{\PP(Y^E_i > Y^C_j + \tau) + 1/2 \, \PP(|Y^E_i - Y^C_j| \leq \tau)}{\PP(Y^C_j > Y^E_i + \tau) + 1/2 \, \PP(|Y^E_i - Y^C_j| \leq \tau)}. \]

The second extension of GPC involves evaluating multiple outcomes to establish a rule for comparing two individuals. For example, consider a patient $i$ in the experimental arm with 27 ventilator-free days but unable to return home at the end of the study. Now, compare this patient to another patient $j$ with 20 ventilator-free days but able to return home. If the comparison focuses solely on ventilator-free days, patient $i$ would be considered to have a more favorable outcome. However, if we consider multiple outcomes simultaneously, one might prioritize the ability to return home over the number of ventilator-free days. In this case, patient $j$ would be considered to have the more desirable outcome.

More formally, let $\bY_i^\ell = (Y_{1i}^\ell, \ldots, Y_{di}^\ell)^\prime$, for $i=1,\ldots,n_\ell$ and $\ellin$, represent the i.i.d. $d$-variate ($d\geq 2$) outcomes of interest. GPC enables the definition of a multivariate rule of comparison, where the scores assigned to each pair reflect an \textit{overall} contrast. In this case, we write $U_{ij}^{\bm{\tau}} = \left [\bY_i^E \succ_{\bm{\tau}} \bY_j^C \right ] - \left [\bY_i^E \prec_{\bm{\tau}} \bY_j^C \right ]$, using bold notations to denote multivariate contrasts. The notation `$\succ_{\bm{\tau}}$' is used to define a more favorable global outcome (with bold subscript $\bm{\tau}$), based on a rule of comparison that incorporates all outcomes in $\bY^E_i$ and $\bY^C_j$. The NTB is then defined in this multivariate setting as $\NTB := \PP(\bY_i^E \succ_{\bm{\tau}} \bY_j^C) - \PP(\bY_i^E \prec_{\bm{\tau}} \bY_j^C)$, using a bold subscript $\bm{\tau}$ as well. 

\subsubsection*{Applying the method to the guiding example}
We now illustrate how a typical GPC-based analysis would be conducted using the guiding example. Table \ref{tab:my_label} presents the results of such an analysis, incorporating the outcomes of death (binary), ability to return home at the end of the study, and ventilator-free days. The table includes three rows, one for each outcome. The rule of comparison between two individuals is here considered to be strictly prioritized by clinical relevance, as is standard practice: patients are first compared on death, then on the ability to return home, and finally on ventilator-free days.

The first row of the table considers a total of $100 \times 100 = 10000$ pairs. Among these, 7\% (labeled `Proportion Favorable') reflect a benefit for the experimental arm (\textit{i.e.}, the experimental patient survived while the control patient did not), 7\% result in the opposite conclusion (`Proportion Unfavorable'), and 85\% (rounded) have the same outcome for both patients (`Proportion Neutral'). The difference between favorable and unfavorable outcomes yields an estimated effect of 0 (rounded), with the associated 95\% confidence interval and $p$-value calculated using asymptotic methods (see \textit{e.g.}, \citet{ozenneinference}).

\begin{table}[t!]
    \centering
    \fontsize{9}{12}\selectfont
    \begin{tabular}{lccccccc}
        Outcome &
        \makecell[c]{Number \\[-5pt] of Pairs} &
        \makecell[c]{Proportion \\[-5pt] Favorable} &
        \makecell[c]{Proportion \\[-5pt] Unfavorable} &
        \makecell[c]{Proportion \\[-5pt] Neutral} &
        \makecell[c]{Cumulative \\[-5pt] NTB} &
        \makecell[c]{95\% \\[-5pt] Confidence Interval} &
        $p$-value \\
        \hline
        \textit{Death} & 10000 & 0.07 & 0.07 & 0.85 & 0 & [-0.07; 0.07] & 0.50 \\
        \textit{Home} & 8540 & 0.27 & 0.15 & 0.44 & 0.12 & [-0.02; 0.26] & 0.09 \\
        \textit{v-f. days} & 4410 & 0.19 & 0.12 & 0.13 & 0.19 & [0.04; 0.34] & 0.015 \\
        \hline
    \end{tabular}
    \caption{Illustration of a GPC-based analysis for the guiding example. A prioritized rule is used for classifying pairs of observations, using the outcomes death, ability to be at home at the end of the study, and ventilator-free days (`v-f. days').}
    \label{tab:my_label}
\end{table}

The second row evaluates the ability to return home. The number of pairs here is 8540, as this outcome is only relevant when there is no difference in the death outcome (the 85\% neutral comparisons on death). Here, 27\% of the original 10000 pairs favor the experimental arm, while 15\% favor the control arm. This results in an estimated NTB of 12\%, considering both the death and return-home outcomes.

Finally, when pairs show no meaningful difference in the first two outcomes, the contrast is made based on the number of ventilator-free days. A threshold of 3 days is set as the minimum clinically-relevant difference for this outcome. In this case, 19\% of the pairs favor the experimental arm, compared to 12\% for the control arm. Accounting for the previous benefits, this leads to an overall cumulative NTB of 0.19.

This example highlights some of the appealing features of GPC, including the accumulation of evidence supporting the benefit of the experimental treatment across various outcomes (other than death in this case) and the ability to identify the specific contributions of each outcome to the overall analysis. In addition to these strengths, GPC's success can be attributed to its apparent simplicity: comparing treatment arms is essentially reduced to the task of contrasting individuals pair by pair. This offers a straightforward framework for integrating clinical relevance and patient preferences into statistical analysis.

However, GPC is not without its challenges. One such challenge stems from the fact that the procedure usually starts with the estimation mechanism, rather than first considering the treatment effect to be estimated. This inherently leads to difficulties in interpretation, especially with nuisances in the data (missing data, censoring), and reflects the `pre-estimand' characteristic of the methodology. Additionally, GPC-based analyses primarily result in claims about (joint) distributions\footnote{For univariate situations, GPC-based effects are sometimes referred to as `overlap measures', as they are measuring the degree of overlap between two densities (which can thus be influenced by the narrowness of inclusion criteria, for instance).}, which, in the context of randomized trials, can be influenced by a variety of factors (see \citet{senn2011u} and \citet{burzykowskilimitations}). Another issue lies in defining the comparison rule. To illustrate, consider two patients: patient $i$, who was able to return home but only had one ventilator-free day, and patient $j$, who was not discharged home but never required a ventilator. According to the prioritized rule, patient $i$ should be considered to have had a more favorable global outcome. However, one might argue that the difference in ventilator-free days is significant enough that it cannot be overlooked when comparing the two patients, regardless of whether or not they went home. In summary, prioritized rules across the outcomes of interest may be overly simplistic to capture clinical relevance. Conversely, developing alternative rules is complex and rarely done, which is why GPC is typically associated with prioritized rules only. This element will be further explored in Section \ref{sec:philo_diff}, after introducing the other methodologies of interest.

\subsection{Desirability of outcome ranking}
The DOOR methodology adopts a different approach to analyzing the overall outcome of patients. It involves two main steps. First, a DOOR outcome is defined: this is an ordinal variable where each category represents mutually exclusive global patient responses, constructed upon several important clinical outcomes. By reducing the multidimensional nature of patient experiences to a single category, this new outcome can then be analyzed in the second step using standard methods for ordinal variables. In spirit, this approach aligns with the procedures used for established ordinal outcomes, such as the the EDSS in multiple sclerosis or the mRS in stroke. The practical difference is that the DOOR literature typically favors non‑parametric methods of analysis (as described below), whereas analyses of the mRS often use either (unadvisable) dichotomization or proportional‑odds models for so-called `shift' analyses. 

More formally, the DOOR methodology maps the original outcomes $\bY_i^\ell$ to a univariate ordinal outcome $\widetilde{Y}_i^\ell$ for $i=1, \dots, n_\ell$, $\ellin$. We arbitrarily consider here that larger values on the ordinal scale represent more desirable experiences. The ordinal variable $\widetilde{Y}$ is then compared across treatment arms, which is typically done via the WMW test. In the context of this paper, the related treatment effects are defined as follows:
\[
\widetilde\Delta := \PP(\widetilde{Y}^E_i > \widetilde{Y}^C_j) - \PP(\widetilde{Y}^E_i < \widetilde{Y}^C_j), \qquad \widetilde\Theta := \frac{\PP(\widetilde{Y}^E_i > \widetilde{Y}^C_j) + 1/2 \, \PP(\widetilde{Y}^E_i = \widetilde{Y}^C_j)}{\PP(\widetilde{Y}^E_i < \widetilde{Y}^C_j) + 1/2 \, \PP(\widetilde{Y}^E_i = \widetilde{Y}^C_j)},
\]
quantifying the contrasts between the \textit{overall} experiences of a randomly-selected patient on the experimental treatment versus a randomly-selected patient on the control. Of note, the DOOR literature usually builds on the so-called DOOR probability $\PP(\widetilde{Y}^E_i > \widetilde{Y}^C_j) + 1/2 \, \PP(\widetilde{Y}^E_i = \widetilde{Y}^C_j)$, which is a  transformation of both $\widetilde\Delta$ and $\widetilde\Theta$.

At this stage, we note that both GPC and DOOR utilize the WMW statistic. However, they do so for different reasons. While GPC's approach is fundamentally built upon the WMW framework, DOOR uses it more as a convenient option rather than a core component. In fact, DOOR could employ another method for analyzing ordinal variables without altering its underlying philosophy. This highlights that the two methods are built on different paradigms, as will be explored later.

One could argue that GPC and DOOR share some conceptual overlap, since any ordinal variable can be represented as a hierarchy of sequentially dichotomized variables. These binary components could then be analyzed using a GPC-type of approach rather than applying DOOR directly to the original ordinal variable. However, we choose not to adopt this perspective, as it offers limited conceptual insight and deviates from the typical spirit of GPC, which is to integrate original outcome variables (or modifications thereof) in the analysis rather than rely on cross-outcome constructions.

\subsubsection*{Applying the method to the guiding example}
Consider the outcomes of the guiding example, with $Y_1$ being the indicator of death (value 1 in case of death), $Y_2$ being the indicator of being at home at the end of the 28-day study, and $Y_3$ being the number of ventilator-free days. 
An illustration of a DOOR outcome is then provided in Table \ref{table:DOOR_outcome_illustrative_example}. 

\begin{table}[t!]
\centering
\fontsize{11}{12}\selectfont 
\begin{tabular}{cll}
$\widetilde{Y}$ & Description (for a 28-day study)& Initial outcomes \\ \hline
1  & Dead & \( Y_1 = 1 \) \\
2  & Alive, not returning home, v-f. days less than 10 & \( Y_1 = 0, Y_2 = 0, Y_3 < 10 \) \\ 
3  & Alive, not returning home, v-f. days between 10 and 25 & \( Y_1 = 0, Y_2 = 0, 10 \leq Y_3 \leq 25 \) \\ 
4  & Alive, returning home, v-f. days less than 10 & \( Y_1 = 0, Y_2 = 1, Y_3 < 10 \) \\ 
5  & Alive, not returning home, v-f. days more than 25 & \( Y_1 = 0, Y_2 = 0, Y_3 > 25 \) \\ 
6  & Alive, returning home, v-f. days between 10 and 25 & \( Y_1 = 0, Y_2 = 1, 10 \leq Y_3 \leq 25 \) \\ 
7  & Alive, returning home, v-f. days more than 25 & \( Y_1 = 0, Y_2 = 1, Y_3 > 25 \) \\ 
\end{tabular}
\caption{DOOR outcome based on the illustrative example - descriptions and corresponding notations. `v-f.' stands for `ventilator-free'.}
\label{table:DOOR_outcome_illustrative_example}
\end{table}

In essence, the definition of the DOOR outcome involves building a partition of patient experiences based on the outcomes of interest $Y_1, Y_2, Y_3$. A ranking of the elements of this partition is then agreed upon, resulting in the ordered values for the DOOR outcome $\widetilde{Y}$. The definition of the DOOR outcome tailored to our example is of course arbitrary, but it contains several points of interest that will be detailed later.

\begin{figure}[t!]
\centering
\includegraphics[width=0.8\linewidth]{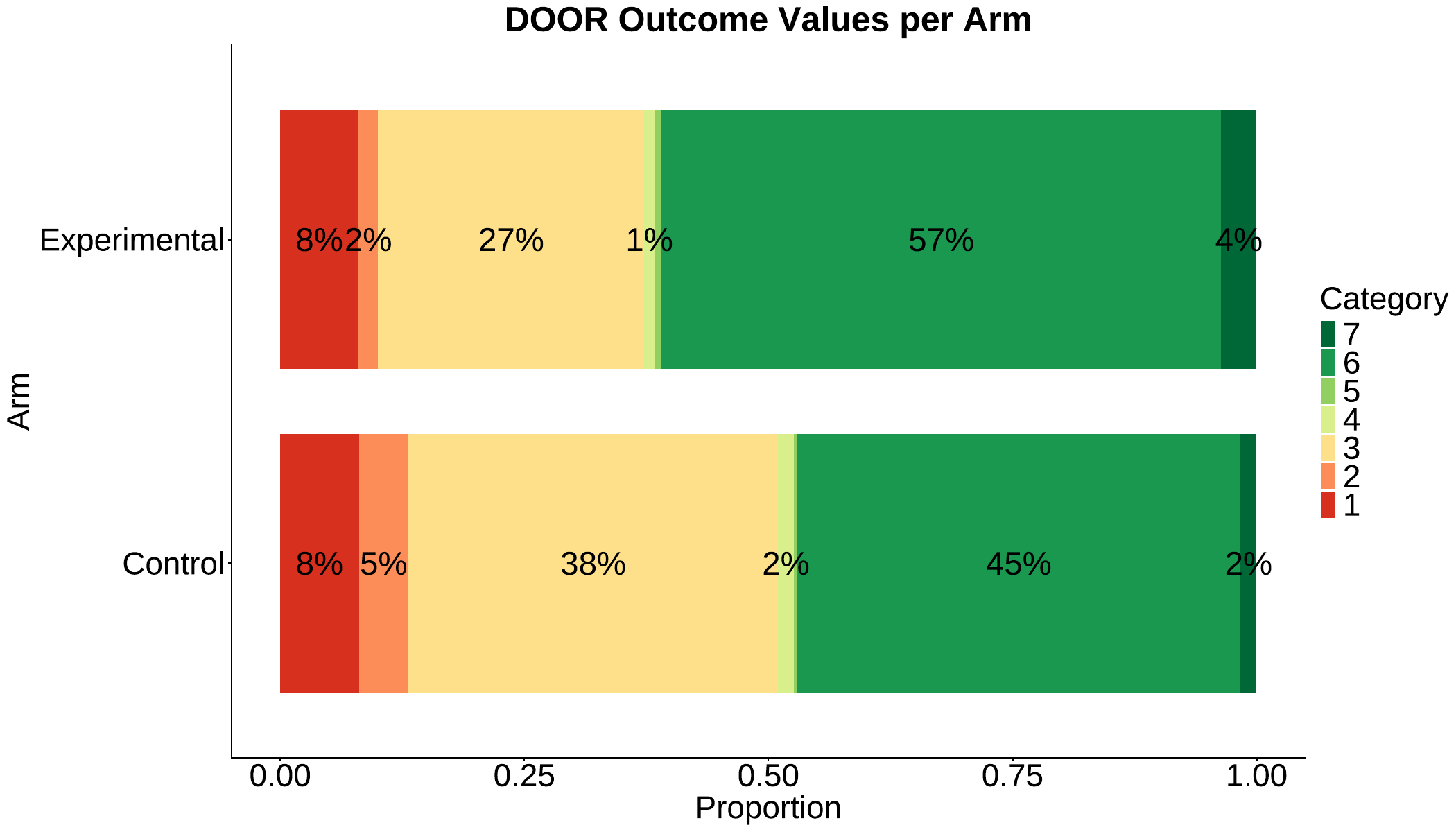}
\caption{\label{fig:DOOR_example} Guiding example: empirical proportions of patients in each of the categories of the DOOR outcome, per arm.}
\end{figure}

A typical DOOR-based analysis then starts with a standard description of the variable $\widetilde{Y}$ for each arm, as reported in Figure \ref{fig:DOOR_example}. Formalization of the difference between arms can then be done with the WMW test, as shown in Table \ref{tab:DOOR_example}. For brevity, we do not include the typical forest plot of the DOOR probability along its components, illustrated for instance in \citet{Toshimitsuapplications}. Of note, Figure \ref{fig:DOOR_example} highlights a point of contrast between GPC and DOOR: while both arms present an observed death proportion of 8\%, GPC summarizes this in Table \ref{tab:my_label} as 7\% favorable \textit{comparisons} and 7\% unfavorable \textit{comparisons}; these are obtained as the proportions of all pairs where one patient died in one arm, and the patient of the other arm survived throughout the study, \textit{i.e.}, $8\% \times 92\% \approx 7\%$.  

The success of DOOR can be attributed to two key factors: first, the combination of outcomes effectively captures various aspects of the treatment effect simultaneously in a way that is easy to communicate, and second, the mapping to a univariate outcome facilitates straightforward analysis and graphical representation.

However, similar to GPC, DOOR is subject to several challenges which will be explored in Sections \ref{sec:philo_diff} and \ref{sec:stat_diff}. For now, we focus on one particular issue: the influence of time. For example, one might argue that the ranking of scenarios in our example could lack consensus in practice because each scenario involves different time dynamics: it is necessary to weigh in certain scenarios whether short-term effects (such as being ventilator-free) are preferred over long-term outcomes (such as being at home at the end of the study). This issue is addressed in the next subsection.

\subsection{Markov Ordinal State Transition Model}
Many of the concepts underlying MOST are to be credited to Frank Harrell, along with the extensive resources available through \citet{fharrell} and the work of \citet{rohde2024bayesian}. We identify two core components that characterize MOST: i) the (discrete-time) serial evaluation of patient health states, and ii) the direct modeling of complete raw data. Regarding the first, MOST can be viewed as a longitudinal extension of DOOR, establishing a clear connection with the previous section and in line with the work of \citet{shu2024longitudinal}. As for the second, MOST takes a different path than the approach proposed in \citet{shu2024longitudinal}. In this section, we elaborate on both aspects.

\begin{table}[t!]
    \centering
    \fontsize{9}{12}\selectfont
    \begin{tabular}{lcccccc}
        Outcome &
        \makecell[c]{Number \\[-5pt] of Pairs} &
        \makecell[c]{Proportion \\[-5pt] Favorable} &
        \makecell[c]{Proportion \\[-5pt] Neutral} &
        \makecell[c]{DOOR \\[-5pt] Probability} &
        \makecell[c]{95\% \\[-5pt] Confidence Interval} &
        $p$-value \\
        \hline
        $\widetilde{Y}$ & 10000    & 0.39       & 0.37        & 0.58 & [0.499; 0.64] & 0.052 
    \end{tabular}
    \caption{Illustration of a DOOR-based analysis for the guiding example, using the DOOR probability $\PP(\widetilde{Y}^E_i > \widetilde{Y}^C_j) + 1/2 \, \PP(\widetilde{Y}^E_i = \widetilde{Y}^C_j)$ as the treatment effect of interest. Inference is here based on similar tools as for GPC - alternative methods can be found in \citet{Toshimitsuapplications}.}
    \label{tab:DOOR_example}
\end{table}

For the first part, the key observation is the challenge introduced by time: one may need to rank the desirability of mild early events versus severe later events. This issue would not arise in a trial so short that the timing of events could be disregarded, in which case consensus on the desirability of experiences could be easily achieved. For example, if the trial in the guiding example lasted only one day, consensus would be straightforward regarding patient status: at home $\succ$ in hospital $\succ$ on ventilator $\succ$ dead, with `$\succ$' denoting `more desirable.' With this in mind, one may consider a longer trial to be essentially a concatenation of short time intervals, each of which allows for consensus to be reached. In other words, a trial can be broken down into short time intervals (\textit{e.g.}, days), and the patient's status can be measured during each of these intervals. Each assessment is then translated into a value on a consensus-based ordinal scale. The overall patient experience is then simply captured by ordinal states measured over time.

In MOST, patient experiences are thus represented as a series of ordinal numbers instead of a single value, offering much more granularity in the evaluation of overall experiences. This has several impacts, both in terms of clinical relevance and statistical opportunities, as will be discussed in Sections \ref{sec:philo_diff} and \ref{sec:stat_diff}. In terms of notation, MOST builds on the ordinal variables $\dot{Y}_i^\ell(t_j)$, for $i=1, \dots, n_\ell$, $\ellin$, measured at each time point $t_j, j = 0, \ldots, \mathcal{J}$. We use here the notation $\dot{Y}$ instead of $\widetilde{Y}$ to emphasize  the difference in nature between these variables: in our guiding example, $\dot{Y}$ represents an ordinal variable of 4 levels (dead, on ventilator, in hospital, at home), while $\widetilde{Y}$ has 7 levels as described previously. 

In addition to its use of longitudinal data, MOST contains a second fundamental characteristic: it emphasizes modeling the raw data as a core component of the methodology. This stands in contrast to a recent extension of DOOR to longitudinal settings by \citet{shu2024longitudinal}, where a DOOR-based contrast summary is computed separately at each time point, as would be done in a traditional DOOR analysis limited to a single time point. That is, at each time-point $t_j$, one estimates for instance the quantity $\dot\Delta(t_j) := \PP(\dot{Y}^E_i(t_j) > \dot{Y}^C_j(t_j)) - \PP(\dot{Y}^E_i(t_j) < \dot{Y}^C_j(t_j))$, for $j = 1, \ldots, \mathcal{J}$. These time-specific contrasts are then combined into an overall measure of difference between arms using a weighted average of the type $\sum_{j=1}^{\mathcal{J}} w_j \dot\Delta(t_j)$, for some pre-specified choice of weights $w_j$, $j = 1, \ldots, \mathcal{J}$. MOST, on the other hand, adopts a modeling-first approach, taking profit of the full trajectory of patient status, including intra-patient dependencies over time, before summarization. In addition to other statistical advantages, this strategy offers more flexible and clinically-meaningful ways of summarizing treatment effects - an interpretational advantage that will be illustrated in Section \ref{sec:philo_diff}.

Many candidate strategies can then be adopted for the modeling of longitudinal data in MOST, see \textit{e.g.}, \citet{harrell2015modeling}. We build here on discrete-time transition models, for which the dependencies of intra-patient outcomes are modeled by conditioning the current outcome on the previous outcome(s). Specifically, we will consider as in \citet{rohde2024bayesian} a first-order transition model where only the previous outcome is conditioned on. Of note, discrete-time transition models offer various advantages over continuous counterparts: estimands are arguably easier to understand, intervals of missing data on the ordinal scale are easier to handle, and state occupancy probabilities (the building blocks of treatment effect definitions in this context) are easier to compute.

For a first-order discrete-time model, the probability for a subject $i$ to be at a given time $t_j$, $j=1, \ldots,\mathcal{J}$ in a state $y$ or higher, given covariates $\bX_i$ (also including time and treatment allocation, allowing to omit the $E$ and $C$ superscripts) and given the entire history of previous states $\dot{Y}_i(t_{j-1}), \ldots, \dot{Y}_i(t_0)$, is a  function of $y$, $\bX_i$, the time $t_j$, the time gap $t_j - t_{j-1}$ and \textit{only} the previous state $\dot{Y}_i(t_{j-1})$. We write:
\begin{equation} \label{eq:MOST_Markov_first_order}
    \mathbb{P}\left(\dot{Y}_i(t_j) \geq y \big| \bm{X}_i, \dot{Y}_i(t_{j-1}), \ldots, \dot{Y}_i(t_0)\right) = g^{-1}\left(\alpha_y + \bm{X}_i\bm{\beta} + h(\dot{Y}_i(t_{j-1}), t_j, t_j - t_{j-1})\right), 
\end{equation}
where the $\alpha_y$'s and $\bm{\beta}$ are regression parameters, $g$ is a link function (\textit{e.g.}, logit link), and where $h$ is a linear function with extra parameters, allowing flexible modeling of the effect of previous states, their time gap with the current state, as well as natural progression effects (\textit{e.g.}, patients' tendency to gradually improve over time and move away from the worst health states). For simplicity, we consider in \eqref{eq:MOST_Markov_first_order} a proportional odds type of model, as the parameter $\bm{\beta}$ is constant for all levels $y$. This can be relaxed in practice, using so-called partial proportional odds models, see \citet{peterson1990partial} and \citet{agresti2013categorical}. 

While we emphasize that modeling raw data is central to the spirit of MOST, the model specified in \eqref{eq:MOST_Markov_first_order} represents just one possible choice. However, transition models, in general, offer several advantages. These include the flexibility in handling different types of serial correlations, the ability to accommodate absorbing states (\textit{e.g.}, death), and compatibility with standard software for independent observations (provided that conditioning on prior states appropriately captures the dependence structure, intra-patient observations are conditionally independent and prior states can be simply added to the set of covariates).

In terms of defining the treatment effect, MOST offers different possibilities given it is model-based. A natural candidate is the parameter in $\bm{\beta}$ associated to treatment allocation. However, there are several reasons one might consider alternative ways of summarizing the contrast between treatment arms: i) difficulties in interpreting an effect on the log-odds scale (if the logit link is used), ii) inability of this summary to capture potential interactions between treatment and time, and iii) incomplete characterization of the treatment effect as the latter can act on transitioning probabilities between distinct states (captured in the $\bm{\beta}$) but also indirectly through the effect of the previous state $\dot{Y}_i(t_{j-1})$ (\textit{e.g.}, patients at home are more likely to stay at home).

Alternatively, treatment effects can be estimated post model fitting, on the basis of (conditional) state occupancy probabilities (SOPs\footnote{To simplify the presentation, we have here considered that all patients start at baseline in the same state. In the absence of additional baseline information, this implies that conditional SOPs are in some sense unconditional. In practice, different SOPs can be calculated for different baseline characteristics and baseline states, and then marginalized by using weighted averages across all combinations of baseline characteristics and states.}). Conditional SOPs capture the probabilities to be in a given state $y$ at a certain time $t_j$, given baseline information, \textit{i.e.} $\PP(\dot{Y}_i(t_j) = y | \dot{Y}_i(t_0), \bX_i)$. Several definitions of treatment effect can then be constructed on the basis of these SOPs. Among these, time-based effects may be of particular interest given their interpretability. For instance, the `mean time unwell', \textit{i.e.}, the average time spent in a state $y$ or worse, can be derived from SOPs for each arm separately, before contrasting the values. Its interpretation echoes that of the restricted mean survival time. Additionally, breaking down the mean time unwell into the mean time spent in each individual state that defines `unwell' also allows for a detailed decomposition of the treatment effect. This will be illustrated when applying the method to the guiding example. Another example of time-based effect is the `days benefit', the average number of days that the outcome for a subject assigned to the experimental arm is better than the outcome for a subject assigned to control, for patients with similar baseline characteristics. 

\subsubsection*{Applying the method to the guiding example}
We consider the four states of interest and arbitrarily associate the ordered numbers 1 to 4 to them, with 1 being death. For illustration, we will use a simplistic version of \eqref{eq:MOST_Markov_first_order}. Letting $X$ be the treatment allocation (value 1 for experimental treatment, 0 for control), we use:
\begin{align*}
\mathbb{P}\left(\dot{Y}_i(t_j) \geq y \,\big|\, \bm{X}_i, \dot{Y}_i(t_{j-1}), \ldots, \dot{Y}_i(t_0)\right) 
= \text{expit}\Big(&\alpha_y + \beta_1 \left[X_i = 1\right] + \beta_2 \,t_j + \beta_3 \left[X_i = 1\right] t_j\\
&+\gamma_1 [\dot{Y}_i(t_{j-1}) = 2] + \gamma_2 [\dot{Y}_i(t_{j-1}) = 3]\\
&+ \gamma_3 [\dot{Y}_i(t_{j-1}) = 4] \Big), 
\end{align*}
for $y = 2, 3, 4$ (for $y=1$ the probability is 1), where $\text{expit}(x) = 1/(1+\exp(-x))$, and where we use distinct notations to highlight elements related to $\bX$ (the $\beta$'s) and elements related to the function $h$ in \eqref{eq:MOST_Markov_first_order} (the $\gamma$'s). The absorbing nature of state 1 (death) is handled beyond the specification of this model, as described later. 

While we introduce some interaction between treatment and time, the model here is simplistic in different ways, among which: i) it considers a linear effect of time, which is unrealistic in many cases, ii) it does not allow for partial proportional odds with respect to time or outcome categories, and iii) it does not include specific effects of time gaps (influence of previous state waning away with greater time gaps) as we consider information to be available on each day. In practice, one may relax these restrictions, in particular the first two, by respectively incorporating the effect of time through flexible methods such as natural cubic splines (allowing for time-varying transition probabilities) and adding extra terms in the model of the type $\gamma_3 [y = 2] t_j$ (for time). The introduction of partial proportional odds with respect to time is particularly relevant when the ordinal scale includes more than one type of clinical event, as different events can occur in different time frames. At the extreme, it is particularly relevant in the presence of absorbing states, since these can only increase across time the log odds of being in that state while different dynamics may be present for the other states (\citet{schildcrout2022model}). Partial proportional odds may also be considered to allow for the treatment effect to vary over outcome categories (\textit{e.g.}, affecting mortality differently to the other categories).

\begin{figure}[t!]
\centering
\includegraphics[width=0.8\linewidth]{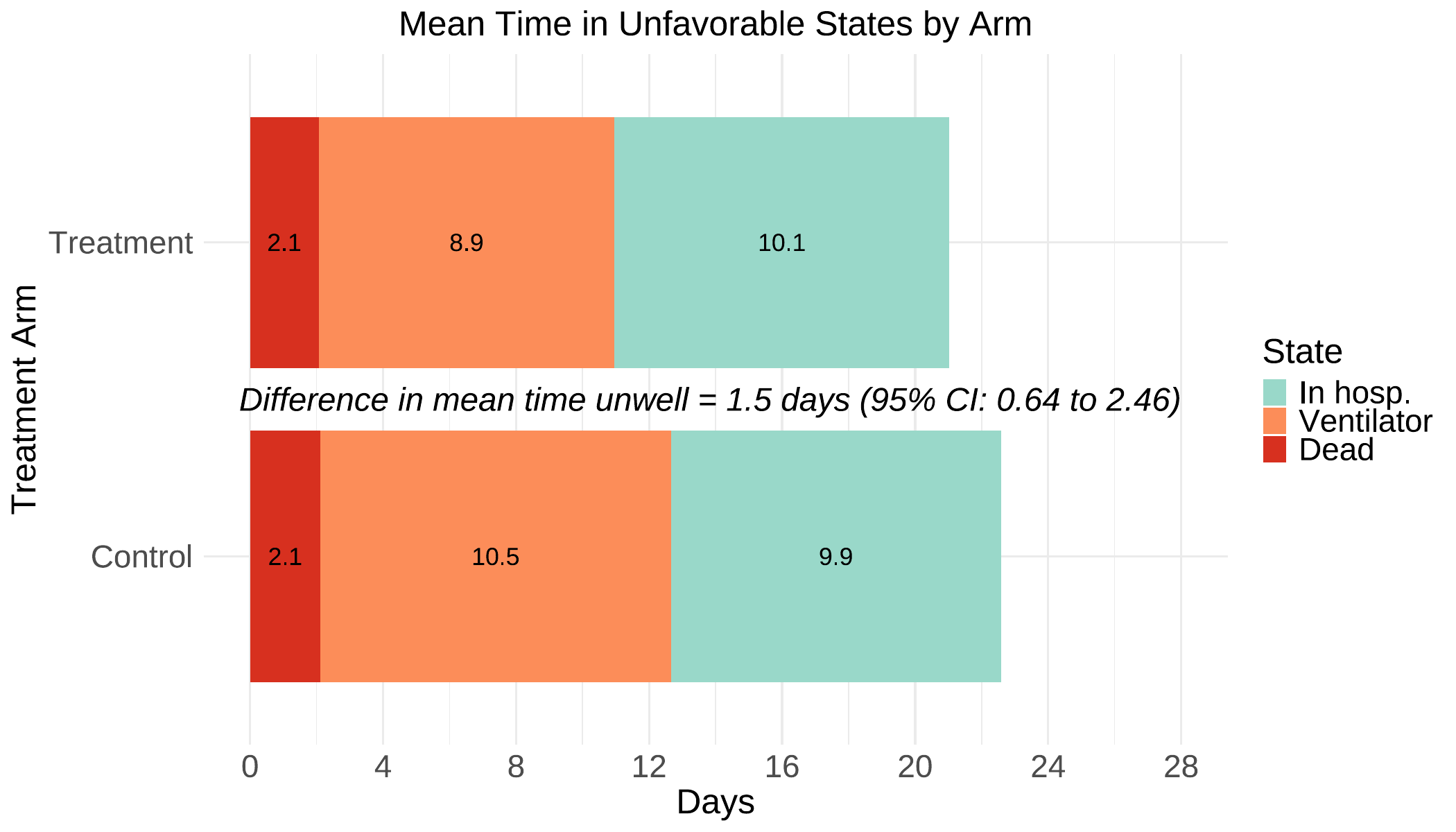}
\caption{\label{fig:MOST_example} Guiding example: mean time unwell (not at home) estimated from the MOST-type of analysis, per arm. Inference is based on a percentile bootstrap approach with 500 replications.}
\end{figure}

We will use here a frequentist approach to the analysis, although Bayesian considerations can be applied (\citet{rohde2024bayesian}). As stated earlier, fitting the model can be done using standard software for independent observations, for instance the function \texttt{rms::orm} in R in the absence of partial proportional odds (\texttt{VGAM::vgam} otherwise). Technical details about model fitting in the presence of absorbing states are given in \citet{schildcrout2022model}. Once fitted, we deduce estimated SOPs for each arm on each day using the method described in \citet{rohde2024bayesian}. In this process, absorbing states are explicitly accounted for by forcing some transition probabilities to be either 1 (indicating permanence in the absorbing state) or 0 (impossible to transition away from the absorbing state). Lastly, we estimate the mean time unwell in each arm, where the notion of `unwell' describes here being in hospital or worse. This is represented in Figure \ref{fig:MOST_example}.

As shown, the experimental treatment yields a benefit in terms of mean time unwell: patients in the experimental arm spend, on average, 1.5 more days at home compared to the control arm. This gain is primarily driven by a reduction in time spent on ventilators: patients in the treated arm required, on average, 1.6 fewer days on ventilator, allowing them to instead remain in hospital or return home. For the sake of brevity, we conclude our application of MOST to the guiding example with this simple contrast. Additional guidance on implementing the method, as well as techniques for verifying underlying assumptions, can be found in \citet{rohde2024bayesian} and \citet{fharrell}. Further considerations around the choice of definition for treatment effect are given in the discussion in Section \ref{sec:discussion}.


\section{Core philosophical differences and their impact} \label{sec:philo_diff}
Through the guiding example, we explore in this section some of the major philosophical differences between the three methodologies. Unless stated otherwise, we use the term DOOR here to denote the non-longitudinal methodology described above, in contrast to its longitudinal version proposed by \citet{shu2024longitudinal}.

\subsection{Contrasting and summarizing}
A first distinction between GPC and both DOOR and MOST lies in the dynamics of contrasting and summarizing. In DOOR, patient experiences are first summarized through the DOOR outcome, and only then are they contrasted across treatment arms. In opposition, GPC is inherently a contrasting methodology. One can think of it as comparing all pairs on \textit{each} individual outcome separately, generating the scores $U_{ij, k}^\tau = \left[Y_{ki}^E \succ_{\tau_k} Y_{kj}^C\right] - \left[Y_{ki}^E \prec_{\tau_k} Y_{kj}^C\right]$, for $k = 1, \ldots, d$ before summarizing these $U_{ij, k}^\tau$'s into a final score $U_{ij}^{\bm{\tau}}$ based on a specific comparison rule. 
  
There is a core implication of this difference: GPC, unlike DOOR and MOST, does not provide any description per arm that is compatible with the contrast of groups. As a result, GPC leaves one with a multivariate evaluation of the difference between arms, without offering insights into how patients performed on each arm separately in a multivariate sense. For example, the analysis of the guiding example suggests a benefit of the experimental treatment, but it cannot provide a quantification of what happened to the patients on  the experimental arm in terms of combinations of death, being at home and ventilator-free days. Instead, one must communicate the result in terms like, ``\textit{For two random patients, one assigned to each treatment, the probability that the patient under the experimental arm will have a more favorable outcome is ...}''. In contrast, DOOR and MOST allow for summarizing the data per arm before making a contrast. This can then be easily communicated, as shown in Figures \ref{fig:DOOR_example} and \ref{fig:MOST_example}. Another graphical tool used by both DOOR and MOST is the so-called `anthology of patient story' plot.

Of note, the ability to estimate how well patients fared separately by each arm is especially important in clinical trials with more than two arms. Without this, methods like GPC-based analyses may fall into issues such as non-transitive conclusions when comparing three or more arms (see, \textit{e.g.}, \citet{thangavelu2007wilcoxon}).


\subsection{Time and continuity}
A second key distinction between the methods lies in how they handle time and continuity, the effects of which are tightly linked.

In both cases, the distinction lies in whether or not categorization is required across time or outcomes. In short, the classical version of DOOR (\textit{i.e.}, the version \textit{not} including a tie-breaker, as will be covered later) will lead to some level of categorization, both across time and across continuous outcomes
. The GPC methodology does not require any categorization, but this comes at a price that will be discussed in this section. MOST aims for a balance, as categorization across time is not required and it can be either exploited or avoided for continuous outcomes depending on the desirability of this action. 

\subsubsection{DOOR}
As mentioned earlier, constructing the DOOR outcome requires partitioning patient experiences. Guidelines for developing this DOOR outcome (available in \citet{hamasaki2025patient}) stress that its categories should reflect clinically meaningful and globally distinct responses. In practice, this may become challenging when the underlying variable is (nearly) continuous. This occurs for instance in the guiding example with ventilator‑free days, where cut‑offs had to be chosen to define the partitions. As with any discretization of  measures, this inevitably leads to information loss. In the guiding example, cutoffs were set at 10 and 25 days for the ventilator-free outcome. Even when these cutoffs are selected on clinical grounds, the issue persists that patients with 9 versus 10 ventilator‑free days will have different contributions to the statistical analysis, while two patients with 10 versus 24 days may contribute similarly.

The use of categorization arises from the challenge of managing the increasing number of scenarios that make up the partition of patient experiences, which must ultimately be ranked by desirability. For example, in epilepsy therapy trials, a consumer-driven approach has been proposed to create a DOOR outcome by incorporating seizure frequencies, adverse events, and quality of life (\citet{vivash2023adaptation}). In this case, categorization is applied both across time and continuous outcomes (for example, quality of life is categorized into `improvement', `no change' and `worsening'). Although this effort aims to compress the information for clarity in ordering, it still results in 60 scenarios to be ranked, due to the factorial combination of categories across outcomes. 

For completeness, we note that some authors have included a continuous outcome in its raw nature in a context of DOOR. This is for instance the case in the DOTS trial (\citet{turner2025dalbavancin}), where quality of life is included as a `tie-breaker'. Another example is the so-called RADAR approach suggested in \citet{evans2015desirability}, which uses days of antibiotic intake as a tie‑breaker for patients who otherwise share the same treatment‑success and adverse‑event overall profile. In both examples however, the tie-breaking nature is more akin to a mix of DOOR and GPC rather than a classical approach to DOOR. For instance, for the DOTS trial, the method with a tie-breaker amounts to defining a GPC-like hierarchy with a 5-level DOOR outcome receiving higher priority over quality of life. The raw nature of quality of life is in that case preserved due to the GPC-part of the approach, such that our comments on categorization of outcomes in the context of more standard DOOR approaches still hold. 

\subsubsection{GPC}
By avoiding the partitioning of patient experiences, GPC does not require any categorization. This allows in some cases for a more nuanced evaluation of the differences between patients by using a scale argument $\tau$: two patients with ventilator-free days values of 9 and 10 may be considered similar for all clinical purposes, while a difference might be declared for values 10 and 24. It is no surprise, then, that GPC and DOOR are applied in different contexts: DOOR is typically used in trials where time-to-event outcomes are not central (\textit{e.g.}, infectious diseases), whereas GPC can be applied in areas where timing is critical, such as overall survival or time to disease progression in oncology.

\begin{figure}[t!]
\centering
\includegraphics[width=0.8\linewidth]{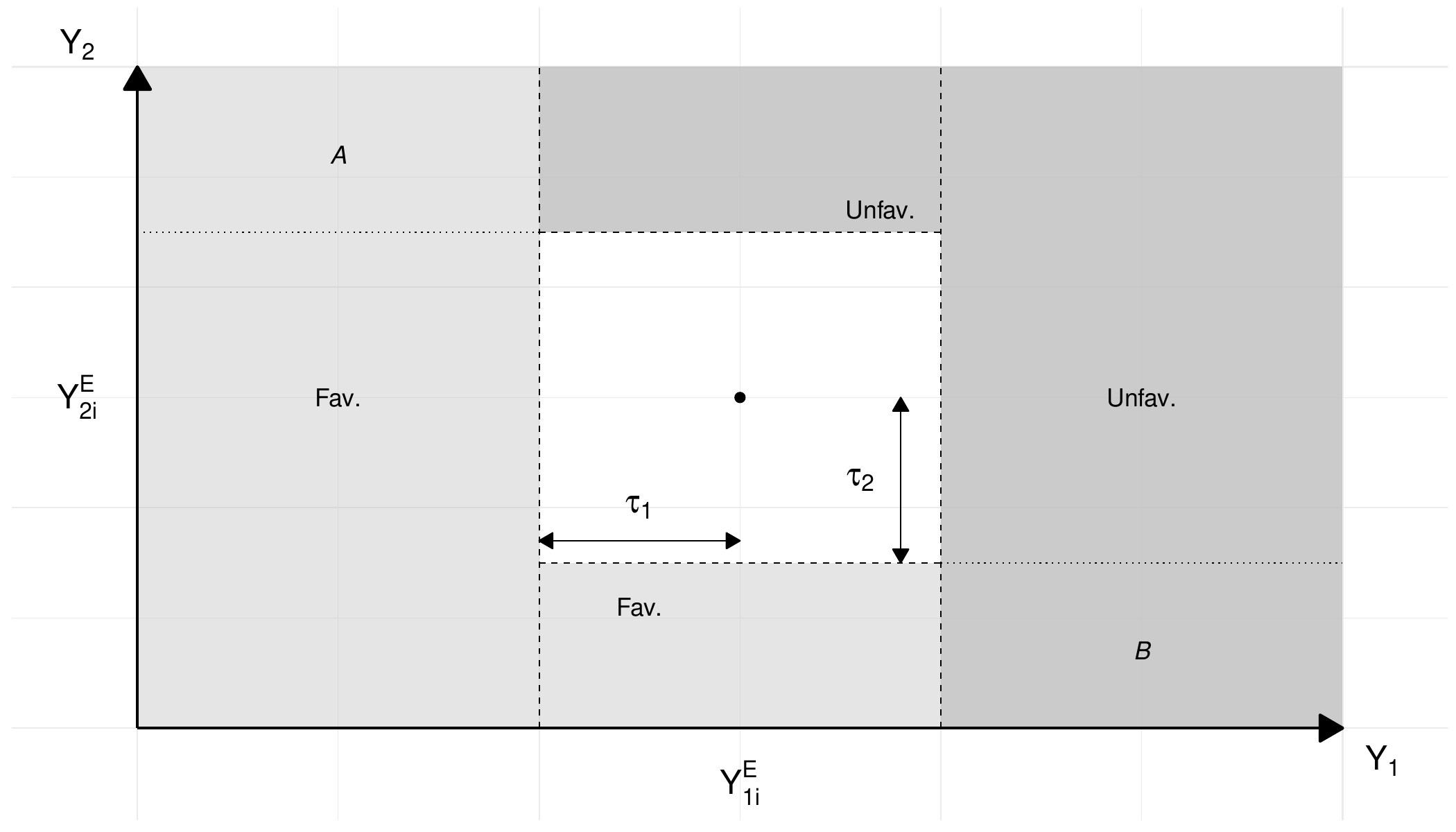}
\caption{\label{fig:GPC_prioritized}Schematic representation of the impact of using prioritized outcomes in GPC. The figure considers two prioritized continuous endpoints $Y_1$ and $Y_2$, assuming higher values are favorable, each subject to a threshold of clinical relevance $\tau_k$, $k=1,2$, under which any difference between patients is considered not to be clinically relevant. The central dot $(Y_{1i}^E, Y_{2i}^E)$ is the data for one patient in the experimental arm. The figure represents the conclusion for the contrast between patient $i$ and any other patient $j$ on control: for instance, if $Y_{1j}^C$ is such that $Y_{1j}^C + \tau_1 < Y_{1i}^E$, then $U_{ij}^{\bm{\tau}} = 1$ and the contrast is in favor (`Fav.' - left part of the figure) of patient $i$, regardless of values for $Y_2$. $A$ and $B$ (upper-left and lower-right) illustrate cases where GPC ignores cross-outcome considerations.}
\end{figure}

There is however a price to pay, coming exactly from not going through the exercise of partitioning global experiences: in practice, (prioritized) GPC might overlook some `cross-outcome' considerations for the evaluation of desirability. In other words, the prioritization of outcomes in GPC comes with a `regardless of what comes after' type of rule: if a difference is observed in the ability to be at home between two patients, this is considered as the definite signal to classify the pair regardless of any information of outcomes having lower priority, in this case ventilator-free days. GPC hereby embodies the concept of ``to choose is to renounce.'' This is conceptualized in Figure \ref{fig:GPC_prioritized}.

Of note, this is purely a consequence of using a prioritized approach to GPC. As stated earlier, GPC, in both its theory and intent, encompasses more than just outcome prioritization. However, creating a classification rule that is fully clinically relevant is a complex undertaking, which has not yet been applied in practice.

The contrast between GPC and DOOR lies thus in balancing the clinical benefits of categorizing certain outcomes (which facilitates cross-outcome considerations) with the drawbacks, both statistical and clinical, of information loss due to arbitrary cutoff values.

\subsubsection{MOST}
By offering to evaluate the status of patients in a succession of small intervals, MOST provides an interesting balance between DOOR and GPC. First, it avoids categorizing over time: inclusion of the (almost exact) timing of events is automatic, provided measurements are done with sufficient granularity (daily, weekly). For instance, in case of univariate binary outcomes capturing terminal events (\textit{e.g.}, death), a MOST-type of analysis (which in this particular case can be seen as a longitudinal binary response analysis) essentially results in the same quantification of the treatment effect as the Cox model (\citet{efron1988logistic}, \citet{fharrellCox}). Additional considerations regarding granularity of measurements across time are developed in Section \ref{sec:stat_diff}.

Like GPC, MOST avoids categorization over time by design, but it further differs by not relying on pairwise comparisons. As a consequence, MOST allows to easily handle differences in follow-up between patients. In contrast, in GPC the comparison of two patients usually accounts for the possibly different lengths of individual follow-up. For example, in cardiovascular studies, comparing only the number of hospitalizations is meaningless if one patient is followed for 6 months and the other one for 3 years. In practice, the comparison of two patients is thus often only considered over a shared duration of follow-up. Besides loss of information, this creates a dependence of the GPC-based test statistic on the distribution of nuisance parameters, in this case the censoring distribution. This is exactly the same issue as that for univariate time-to-event outcomes when resorting to the Gehan-Wilcoxon test (\citet{gehan1965generalized}). In contrast, in MOST, individuals contribute to the analysis on their own, not via comparisons. Having different lengths of follow-up merely implies having different amounts of information per patient, which is automatically accounted for in the likelihood-based method of analysis under certain assumptions.

In MOST, the complexity is reduced in such a way that time-related trade-offs (\textit{e.g.}, weighing early mild events against later severe ones) no longer apply, in contrast to DOOR. This simplification automatically reduces the number of elements that need to be ranked by order of desirability, thereby enabling the inclusion of purely continuous outcomes as well. For example, in cardiovascular trials, a clinical question of interest may initially center around quality of life (QoL), measured on a `continuous' 0-100 scale. However, events like death and stroke are clear clinical overrides that take priority over any value measured for QoL. In this framework, the full response range on a given day might then extend from 0 to 102, with the top two values reserved for the most severe clinical events (supposing lower values on the QoL scale are more desirable).

This creates thus an opportunity in MOST to include continuous outcomes in their raw form. Moreover, the longitudinal nature of MOST eliminates the need to limit inclusion of continuous outcomes to a single time point. This stands in contrast to GPC, where continuous variables are typically only incorporated at a specific time. For example, in the EMPULSE trial (\citet{pocock2023win}) in cardiology, the primary endpoint was a hierarchical GPC-based summary, combining time to death, number of heart failure events, time to first heart failure event, and change in the Kansas City Cardiomyopathy Questionnaire - Total Symptom Score measured \textit{at} day 90. This disregards any information on the latter component measured before day 90. Of note, while GPC is not inherently limited to single-time-point evaluations of continuous outcomes, this remains common practice, largely because integrating longitudinal measures efficiently within GPC frameworks poses significant methodological challenges.

Lastly, in some cases incorporating multiple continuous outcomes at once into MOST can be practically challenging, \textit{e.g.}, considering lab measurements and QoL, both measured on a 0–100 scale. Unless a clear hierarchy of clinical desirability can be considered for the outcomes (hereby creating a 0–200 ordinal scale), it may become necessary to categorize them. This allows the construction of an ordinal scale at each evaluation point that captures clinically meaningful cross-outcome comparisons. This consideration, with both its challenges and advantages, is very similar to that of DOOR, except it is here considered at each time of evaluation, reducing the complexity of the exercise of building the ordinal scale. 

\subsection{Clinical relevance}
A third key point in comparing these methodologies is how they incorporate clinical relevance, which has driven much of the traction behind both GPC and DOOR. For example, in cardiology, GPC steadily replaces time-to-first-event analyses by addressing their limitation of treating events of different types as equally important. By shifting toward a time-to-worst-event approach, GPC helps ensure that early, less clinically-important events don't overshadow more severe events that occur later. This is a welcome advancement, though further opportunities may lie ahead.

A shared limitation of both GPC and DOOR is that clinical relevance tends to be incorporated primarily at the beginning of the process, during the estimation procedure or during the design of the outcome measure, rather than at the end, where the results for the estimand are communicated. For instance, GPC is attractive in interpretability and clinical relevance of the \textit{estimation} procedure, not per se of the underlying estimand; it is a matter of debate whether this estimand, with interpretation in terms of probabilities for two distinct patients, helps practitioners both communicate and quantify what to expect from a particular experimental treatment. Similarly, the clinical relevance of DOOR occurs at the creation of the DOOR outcome, with the same type of challenges for communication around the underlying estimand. The longitudinal version of DOOR suggested by \citet{shu2024longitudinal} is subject to the same challenges, potentially even more pronounced due to the added complexity introduced by weighting. 

Of course, many other measures of treatment effect, commonly used in practice, pose challenges in interpretation, with the hazard ratio being a notable example (\textit{e.g.}, \citet{saad2018understanding} and \citet{verbeeck2024rethinking}). Some of this difficulty may diminish over time as clinicians become more familiar with these metrics. Just as there is now a general sense of what constitutes a `favorable' hazard ratio depending on the clinical context, one could argue that, with time, a similar intuition may develop around what qualifies as a `good' NTB. 

However, this argument is ultimately unconvincing: rather than justifying a limitation by pointing to similar flaws in existing methods, efforts should be directed toward addressing and overcoming the limitation itself. In that sense, MOST may offer a step forward. As discussed earlier, MOST's model-based approach allows for the construction of clinically-meaningful contrasts tailored to the clinical context. In the guiding example, we focused on a so-called mean time unwell, but this is just one possible choice (driven by the consideration that interpretation on a time scale may be easier than other types of summaries). Alternatively, one could define the treatment effect in terms of specific outcome probabilities, such as the probability of being at home at the end of the study (implicitly accounting for death), without having spent an `excessive' amount of time on a ventilator. In essence, MOST gives stakeholders the flexibility to incorporate clinical relevance through two layers rather than one: at the outset, when designing the ordinal scale, and at a second stage of the process, when discussing the (pre-defined) treatment effect that will be used for communication. 

It is important to note the philosophy here: MOST emphasizes maximizing flexibility in modeling before translating the model's output into a treatment effect of interest. The model here is thus only a means to an end. This follows the process of estimating pre-defined quantities of interest \textit{after} analyzing raw data, instead of reducing raw data into quantities of interest. Such an approach is similar in spirit to for instance time-to-event analysis, where survival curves are modeled and then compared at a specific time point, rather than first dichotomizing outcomes (survivors and non-survivors) and applying logistic regression. MOST extends this principle more broadly.

Lastly, MOST naturally accommodates Bayesian perspectives, which introduce new opportunities for interpretation. For example, in the guiding example involving four clinical states, one could jointly consider the probabilities of being in each state under experimental treatment versus control at the end of the study. A treatment effect could then be defined as the posterior probability that \textit{any} of these outcome probabilities are more favorable (by a certain amount, if this nuance is wished for) under experimental treatment. Alternative summary measures are illustrated in \citet{rohde2024bayesian}.



\section{Core statistical differences} \label{sec:stat_diff}

In addition to philosophical differences, the three methodologies also differ in operational and statistical aspects, as detailed below.

\subsection{Nuisances in the data}
The guiding example so far has considered the unrealistic situation of absence of nuisance in the data. In practice, missing data (considered here missing at random) and censoring (non-informative) will both affect the amount of information available. The methods of interest can be contrasted in the way they handle this loss of information. 

Various approaches have been proposed in the literature to address censoring and missing data in the context of GPC and DOOR. For GPC, recent overviews can be found in \citet{buyse2025gpc}. In the case of DOOR, \citet{follmann2020analysis} proposed addressing right-censoring by reformulating it as a form of multiple interval censoring. \citet{shu2025desirability} propose an alternative way of handling survival outcomes in the context of DOOR. To the best of our knowledge, there is limited literature addressing the handling of missing data for DOOR that explicitly account for the outcome's multi-component structure.

For MOST, the impact of censoring due to loss of follow-up is automatically accounted for under assumptions through the likelihood-based approach. A more interesting point is the handling of missing data. Here, the multiple outcomes composing the ordinal scale, as well as the longitudinal aspect of measurements, may help reformulating missing data as interval-censored data (\citet{fharrell}). For example, suppose in the guiding example that a patient's exact status on a particular day is unknown, but it is known that the patient neither passed away nor was at home on that day. In that case, the missing information is transformed into interval censoring in categories $[2, 3]$: the patient's contribution to the likelihood translates presence in one of these two states on that day. For another example, in a QoL study with death as a clinical override, if a QoL measurement is missing at one time but available later, contribution to the likelihood at the missing time will reflect absence of the patient in the state `death' (\textit{i.e.}, left censoring when death is the highest value on the ordinal scale). This will be in contrast to a longitudinal DOOR analysis as in \citet{shu2024longitudinal}, where intra-patient trajectories are not explicitly exploited. The ability to change missing data into censored data in MOST reflects the idea that \textit{some} information is still available on the ordinal scale. This partial information among the outcomes of interest is also used in GPC, but there it is usually exploited through imputation for the missing outcome based on the available information. Of note, this philosophy of viewing missing data as potential partial information in MOST may also apply to DOOR, although this may be more complex given the categorization of DOOR over time. 

\subsection{Information-rich analyses} 
The three methods under consideration share a common principle: they integrate multiple outcomes into a single analytical framework. This approach is often described as information-rich, and is particularly valuable in clinical research settings where small sample sizes present a significant challenge. There is however a difference between the methods, as for a fixed sample size they will benefit differently from the increase in frequency of measuring outcomes.

Because GPC and DOOR (not referring to the work of \citet{shu2024longitudinal}) are non-longitudinal in nature, increasing the frequency of outcome measurements does not lead to substantial efficiency gains. In contrast, MOST can benefit significantly from more frequent measurements, as demonstrated in simulation studies such as \citet{fharrellSim}. In other words, the \textit{exploited} effective sample size (\textit{i.e.}, the sample size one would need in an independent sample to equal the amount of information in the actual longitudinally-measured sample, \citet{faes2009effective}) may be very different between methods.

The extent of this gain is of course context-specific: when outcomes within a patient are weakly dependent, additional measurements contain substantially more novel information than in settings with strong intra-patient dependencies. Similarly, when dependencies decay rapidly over time, later measurements are almost independent from earlier ones, offering information akin to that from an entirely new patient. These are the same types of consideration as in \citet{faes2009effective} for more conventional longitudinal models, where it is shown that under a first-order autoregressive correlation structure, there is theoretically no upper limit to the information gained from additional measurements within a patient. 

MOST offers an additional advantage when it comes to measurement frequency, as increasing the frequency of outcomes considered less important can increase the chance of detection of treatment effects on those outcomes. This, in turn, allows information to be formally borrowed for more critical outcomes that occur less frequently, such as mortality. In the extreme, this corresponds to assuming a proportional odds model, where the treatment effect on mortality is identical to the overall treatment effect. However, and more interestingly, MOST offers here some additional flexibility: such assumptions can be explicitly relaxed and formally modeled. This is done within a Bayesian framework. For instance, one can start by specifying a partial proportional odds model under MOST that allows the treatment effect on mortality to differ from the effect on other outcomes. A prior distribution can then be introduced to reflect the belief that this difference is likely to be small-bounded within a certain factor in either direction. This approach then enables principled borrowing of information while maintaining interpretational clarity. Further discussion on this type of borrowing across outcome levels can be found in \citet{fharrellBorrow}.

\subsection{Intercurrent events}
All three methods under consideration naturally encourage the adoption of the so-called composite strategy for handling intercurrent events, as these events are directly integrated into the outcome definition. For example, in the guiding example, each method explicitly treats death as the highest-priority event.

A distinction between methods will again come from the impact of time, as the information of some intercurrent events can be more effectively addressed by capturing outcomes longitudinally. \citet{simader2024symptoms} provide a nice illustration in cardiology for the incorporation of non-terminal intercurrent events: patients' episodes of angina are recorded daily, and the use of antianginal medication is permitted during the study. The outcome on each day is constructed by combining the number of angina episodes with medication use. For instance, if no medication is taken, the outcome ranges from 0 to 6 based on episode count. If one unit of medication is used, the outcome shifts to a scale of 7 to 13, again reflecting the number of episodes. This pattern continues for increasing medication use. Essentially, this allows the analysis to incorporate the use, the timing and the quantity of concurrent medication. Accounting for this level of granularity is significantly more challenging in methods like GPC and DOOR.

As another example, consider kidney disease trials, where interpreting glomerular filtration rate (GFR) values critically depends on whether they are measured before or after intercurrent events such as dialysis or transplantation. Since time again plays a central role, a MOST-type of approach can be applied by constructing an outcome scale that integrates both GFR measurements and these intercurrent events. On any given day, if neither dialysis nor transplantation ever occurred, the outcome is simply the raw GFR value. However, if either event has taken place, the outcome is assigned a value exceeding the maximum of the raw GFR scale, reflecting a clinically significant change in patient status. In essence, embracing frequent outcome measurement enables a more refined and effective incorporation of intercurrent events into the analysis.

\subsection{Covariate adjustment}

The recent FDA guideline on covariate adjustment (\citet{FDAcov}) emphasizes the value of incorporating prognostic baseline covariates into the design and analysis of clinical trials, while still recognizing the validity of unadjusted analyses. Exploiting baseline prognostic information, whether for marginal or conditional estimands, can generally improve the efficiency of statistical analyses. The underlying mechanisms for this improvement vary depending on the outcome type and modeling strategy, a consideration we do not develop in this paper. However, given that the hierarchical multi-component methods under consideration are often highlighted for their potential to increase statistical power without compromising clinical relevance, it is reasonable to examine how they capitalize on the additional gains in power that covariate adjustment can offer.

In terms of flexibility for incorporating covariates, GPC lags behind both DOOR and MOST. While the DOOR literature generally leans toward non‑parametric methods, both DOOR and MOST may benefit from the well-established body of work on regression modeling for (longitudinal) ordinal outcomes (\textit{e.g.}, \citet{agresti2013categorical}, \citet{neuhaus1992statistical}, \citet{lee2007class}, \citet{follmann2020analysis}). In contrast, regression modeling within the GPC framework remains an area of ongoing methodological development (\citet{thas2025covariate}, \citet{de2025stratification}). While certain approaches have been proposed, they often come with limitations in flexibility: an inability to separate favorable, unfavorable, and neutral results, or to break down overall GPC-based statistics into contributions from individual outcomes.

Moreover, there is a concern among some authors (\citet{thas2025covariate}) about the existence of a pertinent approach to regression modeling for GPC: it is unlikely that realistic data-generating mechanisms for multivariate data are compatible with flexible GPC-based regression models, where the latter would encompass so-called probabilistic index models (\citet{thas2012probabilistic}). This is due to the multivariate nature of GPC, and its ambition to allow for scale arguments (the parameter $\tau$), making regression modeling very context-specific and unlikely to yield a broadly applicable framework. This concern is particularly true if interest lies in conditional estimands, less so for marginal estimands. Consequently, in practical applications, stratification is often used to adjust GPC-based analyses, a method that brings its own set of limitations such as the need to categorize continuous covariates and the resulting constraint on the number of covariates that can be included. As a result, GPC-based analyses may currently be less flexible and efficient than DOOR and MOST when it comes to incorporating covariate information, although numerical investigations are required to supplement this statement.


\section{Areas of further research}\label{sec:further_research}

While this paper does not intend to draw overarching and definite conclusions about the three methodologies, our discussion thus far has highlighted MOST as exhibiting the most desirable properties among them. However, MOST is not without its challenges, and further methodological development is needed. In this section, we outline several avenues for future research, both specific to MOST and applicable to the broader class of hierarchical multi-component methods.

One notable challenge for MOST lies in handling multiple outcomes with differing temporal dynamics within the construction of the ordinal outcome scale. In many clinical settings, some outcomes are measured more frequently than others. For example, in oncology, disease progression is typically assessed at pre-specified times of evaluations, whereas QoL or the experience of toxicities can often be monitored more frequently. Similar patterns arise in other domains, such as epilepsy research, where cognitive assessments may occur infrequently while seizure counts can be recorded daily.

One possible way to address this challenge is to adopt learnings from the handling of missing data. In this framework, the absence of a component outcome on a given day is treated as partial information on the composite ordinal scale. For instance, consider an ordinal scale in oncology that includes QoL on a 0–100 range with higher values being less desirable and with clinical overrides for progression and death at values 101 and 102, respectively. Suppose a patient records a QoL score of 75 on day $t$, and it is known that progression will be confirmed at the next scheduled assessment, while no progression was observed at the previous one occurring before day $t$. In this case, the outcome for day $t$ reflects partial information, as it will be one of the two values $\{75, 101\}$, rather than being a single, definitive value. This type of partial information requires further work, both in methodology and software availability.

There are other directions of research we would like to promote, although our exposition is not exhaustive. For instance, in rare diseases, it may be challenging to provide a consensus on ordering of desirability between outcomes given the heterogeneity of symptoms. In that case, it might be interesting to `individualize' the global outcome. This has been suggested in the context of DOOR (although the methodology is in fact more akin to GPC), where patients enrolled in the trial were invited at baseline to rank their \textit{individual} priorities (\citet{lu2022composite}). A contrast between arms then is based on comparing pairs of patients on a minimal set of common priorities between the two individuals composing the pair. This spirit could be extended to all hierarchical multi-component methodologies, as one could for instance define an ordinal outcome where the ordering of values may represent different realities for different patients, pre-specified at baseline by themselves or even by expert opinion. This brings methodological challenges (\textit{e.g.}, how to handle the possibility for two patients to have a different number of values on their personal scale if one of them can simply not experience a clinical event the other one can), as well as interpretational challenges. 

Finally, all three methodologies face inherent challenges when it comes to dynamic decision-making, such as stopping a trial early for futility or efficacy. This difficulty arises from their joint consideration of multiple outcomes, which naturally introduces a time-related dimension to the analysis. Some components of the composite scale may reflect the treatment effect early in the trial, while others may take longer to manifest. For example, if the experimental treatment shows little impact on early-measured outcomes (\textit{e.g.}, disease progression), there is a risk of prematurely stopping the trial even though meaningful effects on later, potentially more critical outcomes (\textit{e.g.}, mortality) have yet to emerge. In essence, the contribution of each individual outcome to the overall analysis is time-dependent, which in turn makes the interpretation of treatment effects inherently time-sensitive across all three methods. This poses a conceptual as well as a technical challenge: it is a challenge to determine how one could meaningfully link the estimated treatment effect at an interim analysis with that at the end of the study in order to exploit conventional tools for dynamic decision-making. In technical terms, it is likely that the dependence structure of the joint distribution of test statistics at different stages of the trial not only depends on the amount of overlapping information (number of patients or number of events), but also on the time-changing relative contributions of individual outcomes to the multivariate analysis. We note that this issue also affects conventional composite endpoints, although it is often overlooked and requires the implicit assumption that the contributions of individual components remain reasonably constant across different decision-making moments.


\section{Discussion}\label{sec:discussion}


In this paper, we have provided an in-depth discussion of three hierarchical multi-component statistical methodologies. At their core, these approaches challenge us, as medical statisticians, to reconsider discussions about how we define and measure outcomes in clinical trials. This is a long-standing issue: the limitations of simpler approaches, such as responder analyses based on dichotomized data, have been well documented, yet these approaches remain common in practice. The value of the methods in this paper lies in their ability to reframe the problem from a new perspective, offering potential improvements in how we capture and interpret clinical information.

A unifying principle across all three methods is the belief that distinct clinical experiences should contribute differently to the statistical analysis. This perspective is worth promoting, as it encourages analyses that are both information-rich and clinically meaningful. However, the methods diverge in how they conceptualize and incorporate the patient's story. DOOR attempts to reduce the full experience to a single ordinal outcome. GPC avoids such summarization but struggles to fully represent the temporal dimension and potential tradeoffs among outcomes. MOST, by contrast, retains the raw longitudinal data, focusing on the clinical states patients occupy over time. In spirit, MOST offers the most faithful representation of patient trajectories, but it is not without its own challenges, both methodological and practical.

An important point is that, although this paper examines methods for analyzing multiple outcomes jointly, such analyses should always be accompanied by (and not replace) univariate evaluations of individual endpoints. This principle, standard for composite endpoints, applies equally to the more advanced approaches discussed here.

One point of potential imbalance in our comparison of methods is the fact that MOST is inherently model-based, which brings both advantages and limitations. On the positive side, modeling allows for an additional layer of clinical interpretation by widening the options for the definition of `treatment effect', estimation of which could then be derived from the fitted model. It can also lead to reduced variability in estimates in contrast to purely non-parametric methods. However, modeling may be argued to be restrictive in certain contexts, as it is unclear how well a model, however flexible it is, may fit complex ordinal data incorporating many aspects of a patient's experience. This reflects the usual bias-variance tradeoff of modeling. There is no universal solution to the bias concern, and further research grounded in real case studies is necessary to determine when and how modeling approaches might fall short in capturing clinical experiences.

That said, the literature on semiparametric ordinal regression models provides strong evidence of their flexibility and robustness. For instance, these models have shown minimal efficiency loss even when compared to correctly-specified parametric counterparts \citep{liu2017modeling}. This suggests that MOST, by building on such models, is well-positioned to maintain both flexibility and efficiency. Moreover, it allows for the formulation of assumptions that can be explicitly stated and verified in practice.

Two limitations of our investigation stem from the concept of ordering outcomes by desirability. First, we consider that the clinical setting allows for \textit{some} ordering in the desirability of what is measured. This may not always be the case, for instance in rare diseases where many aspects of how a patient feels and functions are affected at once. The GPC literature usually proposes to consider non-prioritized rules of comparisons in that case. Extending the spirit of MOST to this setting, for instance by considering modeling each outcome separately while allowing a link between models, is outside the scope of this paper.  

A second limitation of using an ordinal ordering is that, while simpler to agree on than numerical scales, it ignores how much more desirable an event on one outcome is than, say, two events on another outcome. This is influential when component effects move in opposite directions, such as could be the case in benefit-risk scenarios. For instance, if an experimental treatment cures 50\% of patients allowing them to be at home while the remaining 50\% die, and the control treatment leads to 100\% hospitalization with no deaths, an ordinal scale with `at home $\succ$ hospitalization $\succ$ death' would treat both treatments as equivalent. While trivial consensus can be reached on the ordering at the level of the outcome, it is a matter of debate whether this outcome-level ordering results in a consensus about the `equivalence' in desirability between the two treatment options. As a result, in the absence of quantification (using \textit{e.g.}, utility-based approaches, \citet{buhler2023multistate}), which is outside the scope of this paper, the proposed methods are best applied in settings where component effects are expected to move in the same direction, consistent with standard recommendations for conventional composite endpoints.

Lastly, in this paper we have included the most severe events, such as death, directly in the ordinal scale \textit{and} treatment effect definitions. For example, death was included in all three analyses of the guiding example, including the context-specific definition of mean time unwell. However, this approach may not always be judged appropriate. In some cases, death may indeed warrant a more separate evaluation, especially when the context is such that the timing of death is not clinically relevant. For example, one could argue in the guiding example that the occurrence of death matters in the contrast of arms, but not when it occurs. 
In such situations, a joint approach may be preferable: assessing non-inferiority (or superiority, depending on the context) on mortality and at the same time superiority on the other outcomes combined in a sensible manner (\textit{e.g.}, through the mean time unwell, with `unwell' being on ventilator or hospitalized). Evident statistical gains will then be obtained for this joint evaluation if one uses first a single modeling framework for the experiences of patients, before extracting the two contrasts of interest. For instance, in a frequentist framework, this would translate into power gains for the intersection hypothesis of a closed-testing procedure, due to correlations between test statistics arising from the same underlying fitted model. Further considerations into that direction are outside the scope of this paper.

In summary, all three methods aim to bring statistical analysis closer to the complexity and nuance of patient experiences. Though more sophisticated than traditional approaches, they demonstrate that richer analyses can preserve, and often enhance, clinical interpretability. Embracing this direction is essential if randomized controlled trials are to produce the kind of evidence truly needed to inform effective medical decision-making.


\section*{\small Declarations - Competing Interests}
{ \footnotesize

MDB and VL are full-time employees of UCB.\newline JV is funded by the EU public-private partnership funding health research and innovation project RealiseD, which is supported by the Innovative Health Initiative Joint Undertaking (IHI JU) under grant agreement No 101165912. The JU receives support from the European Union’s Horizon Europe research and innovation programme, as well as from COCIR, EFPIA, Europa Bío, MedTech Europe, and Vaccines Europe.\newline MB is a stockholder of IDDI and One2Treat. \newline MV is a full-time employee of Bayer.\newline SE is a full time employee of George Washington University and reported grants from the National Institute of Allergy and Infectious Diseases (NIAID), National Heart, Lung, and Blood Institute (NHLBI), National Cancer Institute (NCI) and National Institute of Child Health and Human Development (NICHD) of the National Institutes of Health (NIH); the Centers for Disease Control and Prevention; and the Henry Jackson Foundation.\newline TH has no competing interests to declare. \newline FH is a consultant to Baylor Scott \& White Research Institute, Regeneron, and Bayer. \newline The authors declare that they have no other competing interests.

}

\newpage
\bibliographystyle{abbrvnat}
\bibliography{sample}

@article{buyse2010generalized,
  title={Generalized pairwise comparisons of prioritized outcomes in the two-sample problem},
  author={Buyse, Marc},
  journal={Statistics in {M}edicine},
  volume={29},
  number={30},
  pages={3245--3257},
  year={2010},
  publisher={Wiley Online Library}
}

@article{pocock2012win,
  title={The win ratio: a new approach to the analysis of composite endpoints in clinical trials based on clinical priorities},
  author={Pocock, Stuart J and Ariti, Cono A and Collier, Timothy J and Wang, Duolao},
  journal={European {H}eart {J}ournal},
  volume={33},
  number={2},
  pages={176--182},
  year={2012},
  publisher={Oxford University Press}
}

@article{evans2015desirability,
  title={Desirability of outcome ranking ({DOOR}) and response adjusted for duration of antibiotic risk ({RADAR})},
  author={Evans, Scott R and Rubin, Daniel and Follmann, Dean and Pennello, Gene and Huskins, W Charles and Powers, John H and Schoenfeld, David and Chuang-Stein, Christy and Cosgrove, Sara E and Fowler Jr, Vance G and others},
  journal={Clinical {I}nfectious {D}iseases},
  volume={61},
  number={5},
  pages={800--806},
  year={2015},
  publisher={Oxford University Press}
}

@article{pocock2024win,
  title={The win ratio in cardiology trials: lessons learnt, new developments, and wise future use},
  author={Pocock, Stuart J and Gregson, John and Collier, Timothy J and Ferreira, Joao Pedro and Stone, Gregg W},
  journal={European {H}eart {J}ournal},
  volume={45},
  number={44},
  pages={4684--4699},
  year={2024},
  publisher={Oxford University Press UK}
}

@incollection{saadapplications,
  title={Applications in {O}ncology},
  author={Saad, Everardo D and P{\'e}ron, Julien},
  booktitle={Handbook of Generalized Pairwise Comparisons: Methods for Patient-Centric Analysis},
  year={2025},
  pages={365--378},
  publisher={Chapman and Hall/CRC}
}

@incollection{deltuvaiteapplications,
  title={Applications in {R}are {D}iseases},
  author={Deltuvaite-Thomas, Vaiva and Verbeeck, Johan},
  booktitle={Handbook of Generalized Pairwise Comparisons: Methods for Patient-Centric Analysis},
  pages={379--390},
  year={2025},
  publisher={Chapman and Hall/CRC}
}

@incollection{Toshimitsuapplications,
  title={The {D}esirability of {O}utcome {R}anking ({DOOR})},
  author={Toshimitsu, Hamasaki and Evans, Scott},
  booktitle={Handbook of Generalized Pairwise Comparisons: Methods for Patient-Centric Analysis},
  pages={296--312},
  year={2025},
  publisher={Chapman and Hall/CRC}
}

@article{backer2024design,
  title={Design of a clinical trial using generalized pairwise comparisons to test a less intensive treatment regimen},
  author={De Backer, Micka{\"e}l and Sengar, Manju and Mathews, Vikram and Salvaggio, Samuel and Deltuvaite-Thomas, Vaiva and Chi{\^e}m, Jean-Christophe and Saad, Everardo D and Buyse, Marc},
  journal={Clinical {T}rials},
  volume={21},
  number={2},
  pages={180--188},
  year={2024},
  publisher={SAGE Publications Sage UK: London, England}
}

@article{rohde2024bayesian,
  title={Bayesian transition models for ordinal longitudinal outcomes},
  author={Rohde, Maximilian D and French, Benjamin and Stewart, Thomas G and Harrell, Frank},
  journal={Statistics in {M}edicine},
  volume={43},
  number={18},
  pages={3539--3561},
  year={2024},
  publisher={Wiley Online Library}
}

@article{fay2018causal,
  title={Causal estimands and confidence intervals associated with {W}ilcoxon-{M}ann-{W}hitney tests in randomized experiments},
  author={Fay, Michael P and Brittain, Erica H and Shih, Joanna H and Follmann, Dean A and Gabriel, Erin E},
  journal={Statistics in {M}edicine},
  volume={37},
  number={20},
  pages={2923--2937},
  year={2018},
  publisher={Wiley Online Library}
}

@book{buyse2025gpc,
  title={Handbook of Generalized Pairwise Comparisons: Methods for Patient-Centric Analysis},
  author={Buyse, M. and Verbeeck, J. and De Backer, M. and Deltuvaite-Thomas, V. and Saad, E.D. and Molenberghs, G.},
  year={2025},
  publisher={Chapman and Hall/CRC}
}

@article{wilcoxon,
  title={Individual comparisons by ranking methods},
  author={Wilcoxon, Frank},
  journal={Biometrics},
  pages={80--83},
  year={1945},
  volume={1}
}

@article{mann1947test,
  title={On a test of whether one of two random variables is stochastically larger than the other},
  author={Mann, Henry B and Whitney, Donald R},
  journal={The annals of mathematical statistics},
  pages={50--60},
  year={1947},
  publisher={JSTOR}
}

@article{verbeeck2023generalized,
  title={Generalized pairwise comparisons to assess treatment effects: {JACC} review topic of the week},
  author={Verbeeck, Johan and De Backer, Micka{\"e}l and Verwerft, Jan and Salvaggio, Samuel and Valgimigli, Marco and Vranckx, Pascal and Buyse, Marc and Brunner, Edgar},
  journal={Journal of the {A}merican {C}ollege of {C}ardiology},
  volume={82},
  number={13},
  pages={1360--1372},
  year={2023},
  publisher={American College of Cardiology Foundation Washington DC}
}

@incollection{ozenneinference,
  title={Inference},
  author={Ozenne, Brice and Verbeeck, Johan},
  booktitle={Handbook of Generalized Pairwise Comparisons},
  pages={69--97},
  year={2025},
  publisher={Chapman and Hall/CRC}
}

@incollection{buysemolenberghs2025gpc,
  title={{GPC} for {P}atient-{C}entric {T}reatment {D}ecisions},
  author={Buyse, Marc and Molenberghs, Geert},
  booktitle={Handbook of {G}eneralized Pairwise Comparisons},
  pages={3--32},
  year={2025},
  publisher={Chapman and Hall/CRC}
}

@incollection{burzykowskilimitations,
  title={Limitations of the {N}et {T}reatment {B}enefit as a {T}reatment-{E}ffect {M}easure},
  author={Burzykowski, Tomasz},
  booktitle={Handbook of Generalized Pairwise Comparisons},
  pages={151--168},
  year={2025},
  publisher={Chapman and Hall/CRC}
}

@article{senn2011u,
  title={U is for unease: Reasons for mistrusting overlap measures for reporting clinical trials},
  author={Senn, Stephen},
  journal={Statistics in {B}iopharmaceutical {R}esearch},
  volume={3},
  number={2},
  pages={302--309},
  year={2011},
  publisher={Taylor \& Francis}
}

@misc{fharrell,
  author = {Harrell, Frank},
  year = {2025},
  title = {Choice and Construction of Clinical Endpoints},
  howpublished = {\url{https://hbiostat.org/endpoint/}},
  note = {Accessed: 2025-04-01}
}

@article{vivash2023adaptation,
  title={The adaptation of the desirability of outcome ranking for interventional clinical trials in epilepsy: A novel consumer-led outcome measure},
  author={Vivash, Lucy and Johns, Hannah and O'Brien, Terence J and Churilov, Leonid},
  journal={Epilepsia {O}pen},
  volume={8},
  number={4},
  pages={1608--1615},
  year={2023},
  publisher={Wiley Online Library}
}

@article{demets2011historical,
  title={A historical perspective on clinical trials innovation and leadership: where have the academics gone?},
  author={DeMets, David L and Califf, Robert M},
  journal={JAMA},
  volume={305},
  number={7},
  pages={713--714},
  year={2011},
  publisher={American Medical Association}
}

@article{shu2024longitudinal,
  title={Longitudinal benefit: risk analysis through the desirability of outcome ranking ({DOOR}) with application to {ACTT-1} {T}rial},
  author={Shu, Shiyu and Diao, Guoqing and Hamasaki, Toshimitsu and Evans, Scott},
  journal={Statistics in {B}iopharmaceutical {R}esearch},
  pages={1--8},
  year={2024},
  publisher={Taylor \& Francis}
}

@incollection{harrell2015modeling,
  title={Modeling longitudinal responses using generalized least squares},
  author={Harrell, Frank},
  booktitle={Regression Modeling Strategies: With Applications to Linear Models, Logistic and Ordinal Regression, and Survival Analysis},
  pages={143--160},
  year={2015},
  publisher={Springer}
}

@article{little2023validity,
  title={Validity and utility of a hierarchical composite end point for clinical trials of kidney disease progression: a review},
  author={Little, Dustin J and Gasparyan, Samvel B and Schloemer, Patrick and Jongs, Niels and Brinker, Meike and Karpefors, Martin and Tasto, Christoph and Rethemeier, Nicole and Frison, Lars and Nkulikiyinka, Richard and others},
  journal={Journal of the {A}merican {S}ociety of {N}ephrology},
  volume={34},
  number={12},
  pages={1928--1935},
  year={2023},
  publisher={LWW}
}

@article{efron1988logistic,
  title={Logistic regression, survival analysis, and the {K}aplan-{M}eier curve},
  author={Efron, Bradley},
  journal={Journal of the {A}merican {S}tatistical {A}ssociation},
  volume={83},
  number={402},
  pages={414--425},
  year={1988},
  publisher={Taylor \& Francis}
}

@misc{fharrellCox,
  author = {Harrell, Frank},
  year = {2021},
  title = {Longitudinal Binary Logistic Model vs. Cox Model},
  howpublished = {\url{https://hbiostat.org/stat/binarysurv}},
  note = {Accessed: 2025-03-15}
}

@article{peterson1990partial,
  title={Partial proportional odds models for ordinal response variables},
  author={Peterson, Bercedis and Harrell, Frank},
  journal={Journal of the {R}oyal {S}tatistical {S}ociety: {S}eries {C} ({A}pplied {S}tatistics)},
  volume={39},
  number={2},
  pages={205--217},
  year={1990},
  publisher={Wiley Online Library}
}

@book{agresti2013categorical,
  title={Categorical data analysis},
  author={Agresti, Alan},
  year={2013},
  publisher={John Wiley \& Sons}
}

@misc{fharrellSim,
  author = {Harrell, Frank},
  year = {2021},
  title = {Simulating Operating Characteristics of Longitudinal Markov Ordinal Outcome Trials},
  howpublished = {\url{https://hbiostat.org/r/hmisc/markov/sim}},
  note = {Accessed: 2025-02-10}
}

@article{schildcrout2022model,
  title={Model-assisted analyses of longitudinal, ordinal outcomes with absorbing states},
  author={Schildcrout, Jonathan S and Harrell, Frank and Heagerty, Patrick J and Haneuse, Sebastien and Di Gravio, Chiara and Garbett, Shawn P and Rathouz, Paul J and Shepherd, Bryan E},
  journal={Statistics in {M}edicine},
  volume={41},
  number={14},
  pages={2497--2512},
  year={2022},
  publisher={Wiley Online Library}
}

@article{pocock2023win,
  title={The win ratio method in heart failure trials: lessons learnt from {EMPULSE}},
  author={Pocock, Stuart J and Ferreira, Jo{\~a}o Pedro and Collier, Timothy J and Angermann, Christiane E and Biegus, Jan and Collins, Sean P and Kosiborod, Mikhail and Nassif, Michael E and Ponikowski, Piotr and Psotka, Mitchell A and others},
  journal={European {J}ournal of {H}eart {F}ailure},
  volume={25},
  number={5},
  pages={632--641},
  year={2023},
  publisher={Wiley Online Library}
}

@article{follmann2020analysis,
  title={Analysis of ordered composite endpoints},
  author={Follmann, Dean and Fay, Michael P and Hamasaki, Toshimitsu and Evans, Scott},
  journal={Statistics in {M}edicine},
  volume={39},
  number={5},
  pages={602--616},
  year={2020},
  publisher={Wiley Online Library}
}

@article{faes2009effective,
  title={The effective sample size and an alternative small-sample degrees-of-freedom method},
  author={Faes, Christel and Molenberghs, Geert and Aerts, Marc and Verbeke, Geert and Kenward, Michael G},
  journal={The {A}merican {S}tatistician},
  volume={63},
  number={4},
  pages={389--399},
  year={2009},
  publisher={Taylor \& Francis}
}

@misc{fharrellBorrow,
  author = {Harrell, Frank},
  year = {2025},
  title = {Borrowing Information Across Outcomes},
  howpublished = {\url{https://www.fharrell.com/post/yborrow}},
  note = {Accessed: 2025-04-10}
}

@article{simader2024symptoms,
  title={Symptoms as a predictor of the placebo-controlled efficacy of PCI in stable coronary artery disease},
  author={Simader, Florentina A and Rajkumar, Christopher A and Foley, Michael J and Ahmed-Jushuf, Fiyyaz and Chotai, Shayna and Bual, Nina and Khokhar, Arif and Gohar, Aisha and Lampadakis, Ioannis and Ganesananthan, Sashiananthan and others},
  journal={Journal of the {A}merican {C}ollege of {C}ardiology},
  volume={84},
  number={1},
  pages={13--24},
  year={2024},
  publisher={American College of Cardiology Foundation Washington DC}
}

@incollection{thas2025covariate,
  title={Covariate Adjustment for {GPC}},
  author={Thas, Olivier},
  booktitle={Handbook of {G}eneralized {P}airwise {C}omparisons},
  pages={224--248},
  year={2025},
  publisher={Chapman and Hall/CRC}
}

@article{thas2012probabilistic,
  title={Probabilistic index models},
  author={Thas, Olivier and Neve, Jan De and Clement, Lieven and Ottoy, Jean-Pierre},
  journal={Journal of the {R}oyal {S}tatistical {S}ociety {S}eries {B}: {S}tatistical {M}ethodology},
  volume={74},
  number={4},
  pages={623--671},
  year={2012},
  publisher={Oxford University Press}
}

@misc{FDAcov,
  author = {FDA},
  year = {2023},
  title = {Adjusting for Covariates in Randomized Clinical Trials for Drugs and Biological Products},
  howpublished = {\url{https://www.fda.gov/media/148910/download}},
  note = {Accessed: 2025-02-15}
}

@article{gehan1965generalized,
  title={A generalized Wilcoxon test for comparing arbitrarily singly-censored samples},
  author={Gehan, Edmund A},
  journal={Biometrika},
  volume={52},
  number={1-2},
  pages={203--224},
  year={1965},
  publisher={Oxford University Press}
}

@article{lu2022composite,
  title={A composite endpoint for treatment benefit according to patient preference},
  author={Lu, Ying and Zhao, Qian and Zou, Jiying and Yan, Shiyan and Tamaresis, John S and Nelson, Lorene and Tu, Xin M and Chen, Jie and Tian, Lu},
  journal={Statistics in {B}iopharmaceutical {R}esearch},
  volume={14},
  number={4},
  pages={408--422},
  year={2022},
  publisher={Taylor \& Francis}
}

@article{liu2017modeling,
  title={Modeling continuous response variables using ordinal regression},
  author={Liu, Qi and Shepherd, Bryan E and Li, Chun and Harrell, Frank},
  journal={Statistics in {M}edicine},
  volume={36},
  number={27},
  pages={4316--4335},
  year={2017},
  publisher={Wiley Online Library}
}

@article{beigel2020remdesivir,
  author = {John H. Beigel  and Kay M. Tomashek  and Lori E. Dodd  and Aneesh K. Mehta  and Barry S. Zingman  and Andre C. Kalil  and Elizabeth Hohmann  and Helen Y. Chu  and Annie Luetkemeyer  and Susan Kline  and Diego Lopez de Castilla  and Robert W. Finberg  and Kerry Dierberg  and Victor Tapson  and Lanny Hsieh  and Thomas F. Patterson  and Roger Paredes  and Daniel A. Sweeney  and William R. Short  and Giota Touloumi  and David Chien Lye  and Norio Ohmagari  and Myoung-don Oh  and Guillermo M. Ruiz-Palacios  and Thomas Benfield  and Gerd Fätkenheuer  and Mark G. Kortepeter  and Robert L. Atmar  and C. Buddy Creech  and Jens Lundgren  and Abdel G. Babiker  and Sarah Pett  and James D. Neaton  and Timothy H. Burgess  and Tyler Bonnett  and Michelle Green  and Mat Makowski  and Anu Osinusi  and Seema Nayak  and H. Clifford Lane },
title = {Remdesivir for the {T}reatment of {C}ovid-19 — {F}inal {R}eport},
journal = {New {E}ngland {J}ournal of {M}edicine},
volume = {383},
number = {19},
pages = {1813-1826},
year = {2020},
doi = {10.1056/NEJMoa2007764}
}

@misc{ICHE9,
  author       = {{ICH}}, 
  title        = {E9(R1) Addendum on Estimands and Sensitivity Analysis in Clinical Trials to the Guideline on Statistical Principles for Clinical Trials},
  year         = {2019},
  howpublished = {\url{https://database.ich.org/sites/default/files/E9-R1_Step4_Guideline_2019_1203.pdf}},
  note         = {ICH Harmonised Guideline; International Council for Harmonisation of Technical Requirements for Pharmaceuticals for Human Use},
}

@article{neuhaus1992statistical,
  title={Statistical methods for longitudinal and clustered designs with binary responses},
  author={Neuhaus, John M},
  journal={Statistical {M}ethods in {M}edical {R}esearch},
  volume={1},
  number={3},
  pages={249-273},
  year={1992},
  publisher={Sage Publications Sage CA: Thousand Oaks, CA}
}

@article{lee2007class,
  title={A class of {M}arkov models for longitudinal ordinal data},
  author={Lee, Keunbaik and Daniels, Michael J},
  journal={Biometrics},
  volume={63},
  number={4},
  pages={1060-1067},
  year={2007},
  publisher={Oxford University Press}
}

@incollection{de2025stratification,
  title={Stratification and {N}on-{P}arametric {A}djustment},
  author={De Backer, Micka{\"e}l and Fay, Michael},
  booktitle={Handbook of {G}eneralized {P}airwise {C}omparisons},
  pages={187--223},
  year={2025},
  publisher={Chapman and Hall/CRC}
}

@article{buhler2023multistate,
  title={Multistate models as a framework for estimand specification in clinical trials of complex processes},
  author={B{\"u}hler, Alexandra and Cook, Richard J and Lawless, Jerald F},
  journal={Statistics in {M}edicine},
  volume={42},
  number={9},
  pages={1368--1397},
  year={2023},
  publisher={Wiley Online Library}
}

@article{thangavelu2007wilcoxon,
  title={Wilcoxon-{M}ann-{W}hitney test for stratified samples and {E}fron's paradox dice},
  author={Thangavelu, Karthinathan and Brunner, Edgar},
  journal={Journal of {S}tatistical {P}lanning and {I}nference},
  volume={137},
  number={3},
  pages={720-737},
  year={2007},
  publisher={Elsevier}
}

@article{brunner2021win,
  title={Win odds: an adaptation of the win ratio to include ties},
  author={Brunner, Edgar and Vandemeulebroecke, Marc and M{\"u}tze, Tobias},
  journal={Statistics in {M}edicine},
  volume={40},
  number={14},
  pages={3367--3384},
  year={2021},
  publisher={Wiley Online Library}
}

@article{ivanova2016mixed,
  title={Mixed models approaches for joint modeling of different types of responses},
  author={Ivanova, Anna and Molenberghs, Geert and Verbeke, Geert},
  journal={Journal of {B}iopharmaceutical {S}tatistics},
  volume={26},
  number={4},
  pages={601--618},
  year={2016},
  publisher={Taylor \& Francis}
}

@article{iddi2012combined,
  title={A combined overdispersed and marginalized multilevel model},
  author={Iddi, Samuel and Molenberghs, Geert},
  journal={Computational {S}tatistics \& {D}ata {A}nalysis},
  volume={56},
  number={6},
  pages={1944--1951},
  year={2012},
  publisher={Elsevier}
}

@article{saad2018understanding,
  title={Understanding and communicating measures of treatment effect on survival: can we do better?},
  author={Saad, Everardo D and Zalcberg, John R and P{\'e}ron, Julien and Coart, Elisabeth and Burzykowski, Tomasz and Buyse, Marc},
  journal={{JNCI}: {J}ournal of the {N}ational {C}ancer {I}nstitute},
  volume={110},
  number={3},
  pages={232--240},
  year={2018},
  publisher={Oxford University Press}
}

@article{verbeeck2024rethinking,
  title={Rethinking survival analysis: advancing beyond the hazard ratio?},
  author={Verbeeck, Johan and Saad, Everardo D},
  journal={European {H}eart {J}ournal: {A}cute {C}ardiovascular {C}are},
  volume={13},
  number={3},
  pages={313--315},
  year={2024},
  publisher={Oxford University Press US}
}

@article{buyse2021net,
  title={The {N}et {B}enefit of a treatment should take the correlation between benefits and harms into account},
  author={Buyse, Marc and Saad, Everardo D and Peron, Julien and Chiem, Jean-Christophe and De Backer, Micka{\"e}l and Cantagallo, Eva and Ciani, Oriana},
  journal={Journal of {C}linical {E}pidemiology},
  volume={137},
  pages={148--158},
  year={2021},
  publisher={Elsevier}
}

@article{dong2020win,
  title={The win ratio: on interpretation and handling of ties},
  author={Dong, Gaohong and Hoaglin, David C and Qiu, Junshan and Matsouaka, Roland A and Chang, Yu-Wei and Wang, Jiuzhou and Vandemeulebroecke, Marc},
  journal={Statistics in {B}iopharmaceutical {R}esearch},
  year={2020},
  publisher={Taylor \& Francis}
}

@article{evans2022our,
  title={Our most important discovery: the question},
  author={Evans, Scott R},
  journal={Statistics in {B}iopharmaceutical {R}esearch},
  volume={14},
  number={4},
  pages={398--407},
  year={2022},
  publisher={Taylor \& Francis}
}

@article{hamasaki2025patient,
  title={A patient-centric paradigm and tool for clinical research: the {DOOR} is open},
  author={Hamasaki, Toshimitsu and He, Yijie and Wu, Qihang and Howard-Anderson, Jessica and Boucher, Helen W and Doernberg, Sarah B and Holland, Thomas L and Powers III, John H and Wang, Jing and Diao, Guoqing and others},
  journal={Antimicrobial agents and chemotherapy},
  pages={e01478--25},
  year={2025},
  publisher={American Society for Microbiology 1752 N St., NW, Washington, DC}
}

@article{turner2025dalbavancin,
  title={Dalbavancin for treatment of Staphylococcus aureus bacteremia: the {DOTS} randomized clinical trial},
  author={Turner, Nicholas A and Hamasaki, Toshimitsu and Doernberg, Sarah B and Lodise, Thomas P and King, Heather A and Ghazaryan, Varduhi and Cosgrove, Sara E and Jenkins, Timothy C and Liu, Catherine and Sharma, Shrabani and others},
  journal={JAMA},
  volume={334},
  number={10},
  pages={866--877},
  year={2025},
  publisher={American Medical Association}
}

@article{shu2025desirability,
  title={Desirability of outcome ranking ({DOOR}) analysis for multivariate survival outcomes with application to {ACTT}-1 trial},
  author={Shu, Shiyu and Diao, Guoqing and Hamasaki, Toshimitsu and Evans, Scott},
  journal={Clinical {T}rials},
  doi={17407745251385582},
  year={2025},
  publisher={SAGE Publications Sage UK: London, England}
}

@article{verbeeck2025non,
  title={From non-inferiority to superiority: the shift towards patient-centric outcomes},
  author={Verbeeck, Johan and De Backer, Micka{\"e}l and Buyse, Marc},
  journal={European {H}eart {J}ournal: {A}cute {C}ardiovascular {C}are},
  volume={14},
  number={3},
  pages={189--190},
  year={2025},
  publisher={Oxford University Press UK}
}

\end{document}